\documentclass[prd,twocolumn,tightenlines,showpacs,nofootinbib,eqsecnum,amsfonts,amssymb,amsmath]{revtex4}
\usepackage{graphicx}
\begin{document}
%%%%%%%%%%%%%%%%%%%%%%%%%%%%%%%%%%%%%%%%%%%%%%%%%%%%
\newcommand{\dd}{{\rm d}}
\newcommand{\bn}{{\bf n}}
\newcommand{\bx}{{\bf x}}
\newcommand{\bk}{{\bf k}}
\renewcommand{\url}[1]{[{\tt #1}]}
\newcommand{\mat}{{\rm m}}
\newcommand{\de}{{\rm de}}
\newcommand{\etal}{{\em et al.}}
\renewcommand{\:}[2]{{\textstyle\frac{#1}{#2}}}
\renewcommand{\;}[2]{{\frac{#1}{#2}}}
\newcommand{\uudot}{\dot{u}}
\newcommand{\udot}{{\cal A}}
\newcommand{\n}{n}
\newcommand{\N}{N}
\newcommand{\E}{{\cal E}}
\renewcommand{\H}{{\cal H}}
\newcommand{\lc}{\varepsilon}
\newcommand{\hatn}{a}%{{\hat{\n}}}
\newcommand{\dotn}{\alpha}%{{\dot{\n}}}
\newcommand{\lb}{\{}
\newcommand{\rb}{\}}

\newcommand{\bra}[1]{\left(#1\right)}
\newcommand{\bras}[1]{\left[#1\right]}
\newcommand{\brac}[1]{\left\{#1\right\}}
\newcommand{\El}{\mathscr{E}}
\newcommand{\B}{\mathscr{B}}
\newcommand{\nn}{\nonumber}
\newcommand{\curl}{\mathrm{curl}}
\newcommand{\be}{\begin{equation}\fl}
\newcommand{\ee}{\end{equation}}

\newcommand{\bea}{\begin{eqnarray}}
\newcommand{\eea}{\end{eqnarray}}

\newcommand{\sfrac}[2]{{\textstyle{#1\over#2}}}
\renewcommand{\S}{_{\hskip-0.8pt\mathrel{\vcenter{\hbox{\tiny\ooalign
{\raise 1.5pt\hbox{\textsf{S}}}}}}}}
\newcommand{\V}{_{\hskip-0.8pt\mathrel{\vcenter{\hbox{\tiny\ooalign
{\raise 1.5pt\hbox{\textsf{V}}}}}}}}
\newcommand{\T}{_{\hskip-0.8pt\mathrel{\vcenter{\hbox{\tiny\ooalign
{\raise 1.5pt\hbox{\textsf{T}}}}}}}}

\def\cqg{{\em Class. Quantum Grav.\/} }
\def\grg{{\em Gen. Rel. Grav.\/} }
\def\prd{{\em Phys. Rev.\/} {\bf D}}
\def\prl{{\em Phys. Rev. Lett.\/} }
\def\apj{{\em Astrophys. J.\/} }
\def\jmp{{\em J. Math. Phys.\/} }
\def\mn{{\em Mon. Not. Roy. Astr. Soc.\/} }
\def\aph{{\em Ann. Phys. (NY)\/} }
\def\plb{{\em Phys. Lett.\/} {\bf B}}
\def\jmp{{\it J. Math. Phys.}\ }

%%%%%%%%%%%%%%%%%%%%%%%%%%%%%%%%%%%%%%%%%%
\title{How close can an Inhomogeneous Universe mimic the Concordance Model?}
%%%%%%%%%%%%%%%%%%%%%%%%%%%%%%%%%%%%%%%%%%
\author{Peter Dunsby$^{\diamond,\dag,\star}$, Naureen Goheer$^{\dag}$, Bob Osano$^{\dag}$, Jean-Philippe Uzan$^{\dag,\ddag}$}

\affiliation{$\diamond$ Centre for Astrophysics, Cosmology and Gravitation,  University of Cape Town, Rondebosch,
7701, South Africa}

\affiliation{$\dag$ Department of Mathematics and Applied Mathematics, 
University of Cape Town, Rondebosch,
7701, South Africa}

\affiliation{$\star$  South African Astronomical Observatory, 
Observatory, Cape Town, South Africa}

\affiliation{$\ddag$ Institut d'Astrophysique de Paris, UMR-7095 du CNRS, Universit\'e Paris-VI Pierre et Marie
Curie, 98bis bd Arago, F-75014 Paris (France)}

\begin{abstract}
Recently, spatially inhomogeneous cosmological models have been proposed as an alternative to the
$\Lambda$CDM model, with the aim of reproducing the late time dynamics of the Universe
without introducing a cosmological constant or dark energy. This paper investigates the possibility
of distinguishing such models from the standard $\Lambda$CDM using background
or large scale structure data. It also illustrates and emphasizes the necessity of testing the
Copernican principle in order to confront the tests of general relativity with the large scale structure.
\end{abstract}


\date{\today}
\maketitle

%%%%%%%%%%%%%%%%%%%%%%%%%%%%%%%%%%%%%%%%%%%%%%%%%%
\section{Introduction}
%%%%%%%%%%%%%%%%%%%%%%%%%%%%%%%%%%%%%%%%%%%%%%%%%%
The analysis of all existing astrophysical data sets, including the observation of the cosmic microwave background \cite{cmbr}, type Ia supernovae (SNIa) luminosity distance measurements \cite{sneIa} and the large scale structure \cite{lss,baryon,wl}) leads to the  conclusion that our cosmological model requires the introduction of a cosmological constant, representing 72\% of the total matter content of the Universe today,  together with a Cold Dark Matter component which significantly dominates ordinary baryonic matter.  Although such a model remains the best fit to all available cosmological data,
understanding the physical origin of the late time acceleration of the cosmic expansion motivates many investigations on the basis of our cosmological model~\cite{uzanGRG06,uzanCUP}.

As long as the Copernican principle is assumed to hold, there are two possibilities which could explain these observations. Either one introduces a dark energy component (the cosmological constant being the simplest possibility) or one allows for a modification of  General Relativity on astrophysical large scales (see e.g. Ref.~\cite{uzanGRGrev} for a recent review and Ref.~\cite{demodels} for recent
reviews on the large variety of dark energy models).

The standard {\it concordance} model of the universe lies heavily on the assumption that the universe is 
spatially isotropic and homogeneous on large scales and can thus be described 
by a Friedmann-Lema\^{\i}tre (FL) spacetime. Since this principle stands at the heart of our interpretation of all cosmological 
observations, it is important  to ensure that it holds with a sufficient accuracy on the scales on which we perform our observations.  
Recently, various tests of the Copernican principle have been proposed~\cite{uzanPC,stebbinsPC,chrisPC,romanoPC}.
It was also shown that the need for dark energy can be suppressed if our universe is inhomogeneous on scales of some
fraction of the Hubble radius~\cite{ltb1,ltb2,iguchi,ltb3,ltb4}, the underlying reason being that we actually have access
to data only on our past null cone so that it is possible to design spacetime geometries that will enjoy the same (low redshift)
past-light cone structure as the one of a FL model without introducing a cosmological constant~\cite{ltb1,ltb2,iguchi,ltb3}
(see however Ref.~\cite{limitromano} for some limitations).

In any FL model, the complete dynamics of this universe is encoded in a single function of time, the scale factor $a(t)$. 
This implies that all the background observations, e.g., luminosity distance-redshift relation, angular
distance, look-back time, etc.., are functionals of the Hubble rate $H(z)$. More importantly, in a $\Lambda$CDM model,
the late time growth of density perturbation on sub-Hubble scales is also a function of $H(z)$, provided we restrict ourselves to linear evolution. 
This implies that there exist rigidities between different set of independent observables ( from the background and at the level of 
perturbation theory) that can be used to test the underlying hypothesis of  the model~\cite{uzanGRG06,uzanCUP,uzanGRGrev}.

The goal of this article is to investigate how two models that reproduce the same background
data on the past-light cone can be distinguished.  In Section~\ref{sec1}, we recall the main properties of the background spacetime 
and we emphasize that the time drift of the cosmological redshift is  not only in principle 
but also in practice a good test
of the Copernican principle.   Section~\ref{sec2} describes the perturbation theory
and discusses various approximations. In Section~\ref{sec3}, we describe the integration of this simplified model
in order to obtain the transfer functions on the past light-cone. This exercise allows us to 
describe the way one distinguishes a LTB model from a Friedmann model that share the same
light-cone background structure. It also clearly illustrates one of the important limitations of most of
the tests of general relativity on astrophysical scales since they encode the Copernican principle in their construction.

%%%%%%%%%%%%%%%%%%%%%%%%%%%%%%%%%%%%%%%%%%%%%%%%%%
\section{Background spacetime dynamics}\label{sec1}
%%%%%%%%%%%%%%%%%%%%%%%%%%%%%%%%%%%%%%%%%%%%%%%%%%
\subsection{Description of the geometry}
%%%%%%%%%%%%%%%%%%%%%%%%%%%%%%%%%%%%%%%%%%%%%%%%%%
We consider a spatially spherically symmetric and inhomogeneous universe
described by a Lema\^{\i}tre-Tolman-Bondi (LTB) the metric~\cite{ltbmetric}:
\begin{equation}
 \dd s^2 = -\dd t^2 + \frac{X^2(r,t)}{1+2E(r)}{\dd r^2} + R^2(r,t)\dd\Omega^2\,,
\end{equation}
with $X= R'$, using the convention $R'=\partial_rR$ and $\dot R=\partial_t R$.
It is convenient to define
\begin{equation}
 2E(r)\equiv -k(r)r^2,
\end{equation}
as well as the two Hubble expansion rates
\begin{equation}
 H_\perp \equiv \frac{\dot R}{R}, \qquad
 H_\parallel \equiv \frac{\dot X}{X} = \frac{\dot R'}{R'}\,.
\end{equation}
The field equations for such a spacetime are given by
\begin{equation}\label{fleq}
 \frac{\dot R^2}{R^2} = \frac{M(r)}{R^3} + \frac{2E(r)}{R^2},\qquad
 8\pi G\rho(r,t) = \frac{M'}{R^2R'},
\end{equation}
from which it can be checked that the continuity
equation 
\begin{equation}
 \dot\rho + (2H_\perp+H_\parallel)\rho=0
\end{equation}
is satisfied. These equations can be solved parametrically as
\begin{equation}\label{soleq}
 R(t,r) = \frac{m(r) r}{2\hat k(r)} \phi'(\eta)\,,
 \quad
 t-t_B(r) = \frac{m(r)}{2 \hat k^{3/2}(r)} \phi(\eta)\,,
\end{equation}
where we have defined $\phi(\eta)=(\eta-\sin\eta,\eta^3/6,\sinh\eta-\eta)$ 
and $\hat k =(k,r^{-2},-k)$ respectively
for $k$ positive, null and negative and defined $m(r)=M(r)/r^3$.
This solution involves three arbitrary functions $M(r)$, $t_B(r)$
and $k(r)$ (or equivalently $E(r)$) but only two of them are actually needed since
one can fix one of them by a proper choice of
the radial coordinate $r$.

To finish this description of the background spacetime, let us compute
the expressions of $\dot R$, $R'$ and $\dot R'$ that we will need later on.
It is obvious from Eq.~(\ref{fleq}) that
\begin{equation}\label{ltbRdot}
 \dot R = \sqrt{ \frac{M(r)}{R} + 2E(r)}.
\end{equation}
Then, from Eq.~(\ref{soleq}) we deduce that
\begin{equation}\label{ltbRprim}
 R' = \left(\frac{M'}{M} -\frac{E'}{E}\right) R - \left[t_B' - \left( \frac{3}{2}\frac{E'}{E}-\frac{M'}{M}\right)(t-t_B) \right)\dot R,
\end{equation}
and then that
\begin{eqnarray}\label{ltbRdotprim}
 \dot R' &=& \frac{1}{2}\frac{E'}{E}\dot R \\
            &&+\left[t'_B-\left(  \frac{3}{2}\frac{E'}{E}-\frac{M'}{M}\right) (t-t_B)\right]\frac{M}{2R^2},\nonumber\\
   &=& \frac{1}{2\dot R}\left(\frac{M'}{R}-\frac{MR'}{R^2}+2E'\right).
\end{eqnarray}

The case of a spatially homogeneous universe is recovered in the case
where $t_B(r)=0$, $m=$constant and $k=$~constant. It follows that
$R(r,t)=a(t)r$ and $X(r,t)=a(t)$ where $a$ is the scale factor. The
spacetime metric thus takes a FL form
\begin{equation}
 \dd s^2 = -\dd t^2 + a^2(t)\left[\dd\chi^2 + f^2_K(\chi)\dd\Omega^2\right],
\end{equation}
with $f_K(\chi)=(\sin\chi,\chi,\sinh\chi)$ depending on the sign of $K$.

%%%%%%%%%%%%%%%%%%%%%%%%%%%%%%%%%%%%%%%%%%%%%%%%%%
\subsection{Light cone equation}
%%%%%%%%%%%%%%%%%%%%%%%%%%%%%%%%%%%%%%%%%%%%%%%%%%
Most of our analysis focuses on what is actually observed
on our past light-cone, which can be defined by
solving the null geodesic equation.
Given a null geodesic with tangent vector $k^\mu$, the
redshift of any object is defined by
\begin{equation}
 1+z \equiv \frac{(k_\mu u^\mu)_{em}}{(k_\mu u^\mu)_{rec}}\,,
\end{equation}
where $u^\mu$ is the tangent vector to the matter worldlines.
It follows that the geodesic equation $k^\mu\nabla_\mu k^\nu=0$
for a null-vector reduces to
\begin{eqnarray}
 \frac{\dd t}{\dd z} &=&-\frac{1}{(1+z)H_\parallel},\label{geo1}\\
 \frac{\dd r}{\dd z} &=&\frac{\sqrt{1+2E(r)}}{(1+z)\dot R'}.\label{geo2}
\end{eqnarray}
This defines our past lightcone that we shall denote by
$$
 {\mathcal C}_-: \,\, \lbrace r=r_*(z), t=t_*(z), \quad r_*(0)=0, t_*(0)=t_0 \rbrace.
$$
The redshift is the observational radial coordinate which can be 
expressed as a look-back time or distance, assuming a cosmological
model~\cite{stoegeretal}. One can indeed use either $r$, $t$ or $z$ as integration
variable, the important point being that only 2 of the three
quantities $(t_*,r_*,z)$ are independent.

When evaluated along the past light-cone, the function $R(r,t)$ is related to the angular distance $D_A$
by
\begin{equation}
 {\cal R}_0(z) \equiv R[t_*(z),r_*(z)]=D_A(z)\,.
\end{equation}
This observational relation can be used to set one constraints on the two arbitrary functions
of the LTB spacetime.
It is useful to redefine the functions introduced previously when evaluated on the past-light
cone as
\begin{equation}
   {\cal R}_{10}\equiv \dot R[t_*,r_*],\quad
   {\cal R}_{01}\equiv  R'[t_*,r_*],
\end{equation}
and
\begin{equation}
   {\cal R}_{11}\equiv \dot R'[t_*,r_*],
\end{equation}
to be considered either as functions of $z$ or $r_*$

We choose $r_*$ as the radial distance along the light cone
and fix it by imposing
\begin{equation}\label{choicer}
 \frac{{\cal R}_{01}}{\sqrt{1+2E}}=1,
\end{equation}
which simplifies the past light-cone equation reducing it to
\begin{eqnarray}\label{eqx1}
 \frac{\dd t_*}{\dd r_*} = -1, \qquad
 \frac{\dd z}{\dd r_*} = \frac{(1+z){\cal R}_{11}(r_*)}{\sqrt{1+2E(r_*)}}.
\end{eqnarray}

%%%%%%%%%%%%%%%%%%%%%%%%%%%%%%%%%%%%%%%%%%%%%%%%%%
\subsection{Reconstruction procedure of a LTB geometry}\label{subsecIIc}
%%%%%%%%%%%%%%%%%%%%%%%%%%%%%%%%%%%%%%%%%%%%%%%%%%
\subsubsection{General procedure}
%%%%%%%%%%%%%%%%%%%%%%%%%%%%%%%%%%%%%%%%%%%%%%%%%

With the choice of coordinates~(\ref{choicer})  the derivative of ${\cal R}$ along the past light-cone
is given by
\begin{eqnarray}\label{dRdr}
 \frac{\dd {\cal R}}{\dd r_*} &=&  {\cal R}_{01}- {\cal R}_{10}
\end{eqnarray}
and Eqs.~(\ref{ltbRprim}-\ref{ltbRdotprim}) can be rewritten as
\begin{eqnarray}
 &&\left[{\cal R}_0 - (t-t_B){\cal R}_{10}\right]\frac{M'}{M}
 +\left[\frac{3}{2}(t-t_B){\cal R}_{10}-{\cal R}_0\right]\frac{E'}{E}\nonumber\\
 && \qquad\qquad\qquad - {\cal R}_{10} t'_B = {\cal R}_{01}=\sqrt{1+2E},\label{eqx2}\\
 && \frac{M}{2{\cal R}^2_0}\frac{M'}{M}  (t-t_B)+\left[\frac{1}{2}{\cal R}_{10} -\frac{3}{2} (t-t_B)\frac{M}{2{\cal R}_0^2}\right]\frac{E'}{E}
 \nonumber\\
 &&\qquad+\frac{M}{2{\cal R}^2_0} t'_B  ={\cal R}_{11}\label{eqx3}.
\end{eqnarray}

Hence, provided $D_A(z)$ is known from observation, we fix ${\cal R}_0$ and then 
Eqs.~(\ref{eqx1}-\ref{eqx3}) give 4 equations for the 6 unknown functions $(t,z,M,E,t_B,{\cal R})$ of $r_*$.
One thus need to fix 2 conditions to completely specify the model.

This approach was investigated in Ref.~\cite{iguchi} who designed various LTB solution sharing the same
$D_A(z)$ relation as a FL spacetime by imposing either $t_B=0$ or $k=$~constant. We follow
the same line and consider LTB models that reproduce the angular distance-redshift relation of
a fiducial standard flat $\Lambda$CDM model. For such a FL universe
\begin{equation}
 \frac{H_{FL}^2}{H^2_0} = \Omega_{\mat0} (1+z)^3 + \Omega_{\Lambda0},
\end{equation}
with $\Omega_{\Lambda0}=1-\Omega_{\mat0}$. $H_0$ is the
Hubble parameter evaluated today and the density parameters
are defined by $\Omega_{\mat0}=8\pi G\rho_{\mat0}/{3H_0^2}$
and $\Omega_{\Lambda0}={\Lambda}/{3H_0^2}$. 
The angular diameter distance is then given by
\begin{equation}
 D_A(z) = \frac{1}{H_0(1+z)}\int_0^z\frac{\dd u}{\sqrt{\Omega_{\mat0}(1+u)^3+ \Omega_{\Lambda0}}}.
\end{equation}
We thus impose that
\begin{equation}
 {\cal R}(z) = D_A(z)\,,
\end{equation}
so that both cosmological models enjoy the same angular distance-redshift relation
(and thus luminosity distance-redshift relation, as can be observationally
determined from SNIa observations).

%%%%%%%%%%%%%%%%%%%%%%%%%%%%%%%%%%%%%%%%%%%%%%%%%
\subsubsection{A model mimicking the background light-cone dynamics of a FL universe}
%%%%%%%%%%%%%%%%%%%%%%%%%%%%%%%%%%%%%%%%%%%%%%%%%

To completely specify the model, we need to impose another condition on 
the LTB model. In order to construct a model that would mimic  
a fiducial flat $\Lambda$-CDM as close as possible, we follow Ref.~\cite{ltb1}
and further assume that the LTB matter energy density distribution along the light-cone, $\rho(z)$,
matches the observed mass density as a function of redshift as determined
in the fiducial $\Lambda$CDM model, that is
\begin{equation}
 8\pi G\rho(z) = 8\pi G\rho_{FL}(z)=3 \Omega_{\mat0}H_0^2(1+z)^3.
\end{equation}
This assumption allows to determine
$r_*(z)$~\cite{ltb1,ltb3}, and in the particular case of a flat  $\Lambda$CDM,
this implies~\cite{kolb}
\begin{equation}
 \frac{\dd r_*}{\dd z} = \frac{1}{(1+z)H_{FL}(z)}.
\end{equation}
Comparing with Eq.~(\ref{geo2}), this means that, under
the choice~(\ref{choicer}) for the radial coordinate, $H_\parallel(z)=H_{FL}(z)$.

\begin{figure}[htp]
\includegraphics[width=8cm]{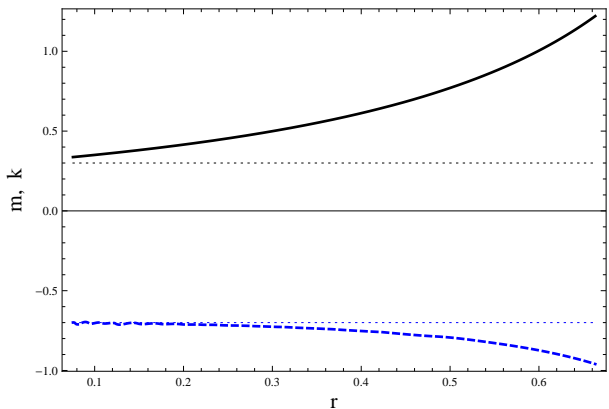}
\caption{Reconstruction of the function $m(r)$ (black, solid line) and $k(r)$
(blue dashed line)  entering the
definition of the LTB geometry for a spacetime reproducing both
$D_A(z)$ and $\rho_\mat(z)$ on the past light-cone. The light dotted lines
correspond to $m(0)$ and $k(0)$.} 
\label{fig0}
\end{figure}

We then follow the reconstruction procedure described in Refs.~\cite{ltb1,ltb3}
and our result agrees with those published in these works. Figure~\ref{fig0} depicts
the function $m(r)$ and $k(r)$ obtained from this reconstruction.

Eqs.~(\ref{ltbRdot}-\ref{ltbRdotprim}) imply that when $r\rightarrow0$, $R/r\rightarrow1$
while $m\rightarrow\Omega_{\mat0}H_0^2$ and $k\rightarrow(\Omega_{\mat0}-1)H_0^2$.
The last free function $t_B(r)$ can be reconstructed from Eq.~(\ref{ltbRdotprim}) but we
will not need it explicitly in the following. Imposing that $t_B(0)=0$, we can compute
$t_0$ from (\ref{soleq}), after evaluating $\eta_0$.

%%%%%%%%%%%%%%%%%%%%%%%%%%%%%%%%%%%%%%%%%%%%%%%%%
\subsection{Distinguishing the two models}
%%%%%%%%%%%%%%%%%%%%%%%%%%%%%%%%%%%%%%%%%%%%%%%%%

The LTB model defined in the previous section was
designed to strictly mimic the luminosity (or angular) distance redshift
relation and the mass density-redshift relation of
the fiducial $\Lambda$CDM model.

In order to distinguish these two models, one needs to
find another independent observable quantity. It
was recently pointed out that one can extract
some information of the dynamics off the light-cone
by considering the time drift of the cosmological
redshifts~\cite{uzanPC}. While advocated as a test
of the Copernican principle its
amplitude in a non-FL model was not
estimated. This can easily be obtained for
the LTB model under investigation. We first
remind that the time drift of cosmological
redshift in a LTB universe takes the form~\cite{uzanPC,chrisrevue}
\begin{equation}
 \dot z = (1+z)H_0 - H_\perp(z),
\end{equation}
which generalized the original FL-expression~\cite{sandage}.
Indeed in a FL-model $H_\perp=H_\parallel=H$ so that $\dot z$
derives from $H(z)$, as any other background observations.
However, in our LTB-model
$H_\perp\not= H_\parallel=H_{FL}$. This was
used to demonstrate that $\dot z[z]$ allows to fully close the reconstruction system
without resorting to making assumption on the matter energy density profile along the light cone.

\begin{figure}[htp]
\includegraphics[width=8cm]{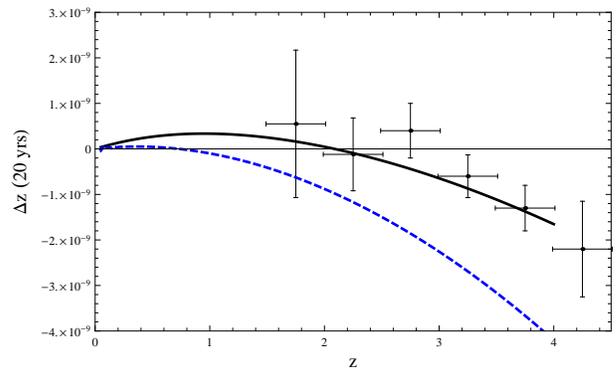}
\caption{Time drift of the cosmological redshift for the standard $\Lambda$-CDM
model (black, solid line) and a LTB-model (blue, dashed line) designed to share
the same observational relation on the past light-cone. The data points
follow the estimates of Ref.~\cite{pasquini} for a CODEX-like spectrograph
on an ELT.} 
\label{fig1}
\end{figure}

Figure~\ref{fig1} compares the expected time drifts of the cosmological
redshift for the LTB- and FL-models. $\Delta z =\dot z\Delta t_{\rm obs}$ has a typical
amplitude of order $10^{-9}$ on a time scale of
$\Delta t_{\rm obs}=20$~yr, for a source at redshift $z\sim4$. 
This measurement is impossible with present-day facilities.
However, it was recently revisited~\cite{loeb} in the
context of ELT, arguing they could measure velocity shifts
of order $1Ð10$ cm/s over a 10~yr period from the
observation of the Lyman-$\alpha$ forest. It is one of the science
drivers in design of the CODEX ultrastable spectrograph~\cite{pasquini} for the
future European ELT. Indeed, many
effects, such as proper motion of the sources, local gravitational
potential, or acceleration of the Sun may contribute
to the time drift of the redshift. It was shown~\cite{ubm,liske},
however, that these contributions can be brought to a
0.1\% level so that the cosmological redshift is actually
measured. The data points and
error bars of Fig.~\ref{fig1} follows the forecast of Ref.~\cite{pasquini}.
Our analysis confirms the recent analysis by Ref.~\cite{quartin},
which also suggests to use the cosmic parallax.

%%%%%%%%%%%%%%%%%%%%%%%%%%%%%%%%%%%%%%%%%%%%%%%%%
\subsection{Discussion}
%%%%%%%%%%%%%%%%%%%%%%%%%%%%%%%%%%%%%%%%%%%%%%%%%

The result of Fig.~\ref{fig1} demonstrates that the information off the light-cone
allows to distinguish a $\Lambda$-CDM from a LTB-model specially designed
to have the same luminosity (or angular) distance redshift
relation and the same mass density-redshift relation as the a $\Lambda$-CDM.
In the case of a dark energy model both $D_L(z)$ and $\dot z$ are
modified, and little insight is gained on the equation of state from adding
the new information on $\dot z$~\cite{corasaniti}. The case of the large scale
inhomogeneity turns out to be different and the $\dot z$ observation
would provide, when available, an interesting extra-test of these models
that cannot be performed otherwise.

Indeed, Fig.~\ref{fig1} was obtained by forcing $D_A(z)$ to match the
$\Lambda$CDM prediction up to $z\simeq4$ while SNIa data~\cite{sneIa}
extend roughly to $z\sim1.6$. Following e.g. Ref.~\cite{february}, one can try to design
density profiles such that the LTB model reproduces the $\Lambda$CDM-$D_A(z)$
at low redshift and becomes homogeneous on large scales. Note that reproducing
$\rho_{\FL}$ and $D_A$ of a flat-$\Lambda$CDM imposes
that $H_\perp(z)=H_{FL}(z)$. Now, assuming that it fits $\dot z_{FL}$ and 
$\rho_{\FL}$ at high redshift implies that
$H_\perp=H_\parallel=H_{FL}$. While attractive, such models seem however difficult
to construct since they usually require that $M'<0$ between these two regimes.

This example demonstrates the complementarity of these two observables since they 
concern two domains of redshift.
To go further in distinguishing such a model from its FL-twin, we shall now
consider the influence of the large scale inhomogeneity 
on the growth of large scale structures.

%%%%%%%%%%%%%%%%%%%%%%%%%%%%%%%%%%%%%%%%%%%
\section{Evolution of density perturbation in a LTB universe}\label{sec2}
%%%%%%%%%%%%%%%%%%%%%%%%%%%%%%%%%%%%%%%%%%%

A general study of the perturbation theory around a LTB background was
performed using a coordinate based approach in Ref.~\cite{clarkson}, and
the general features of the growth of density perturbations were discussed
in Ref.~\cite{zibin}.  The goal of this section is to investigate the evolution of the density contrast using
the 1+1+2 formalism, and to obtain an approximation for the evolution equations.
We also discuss how the density contrast variable introduced here can be
related to observations.

%%%%%%%%%%%%%%%%%%%%%%%%%%%%%%%%%%%%%%%%%%%%%%%%%
\subsection{General formalism}
%%%%%%%%%%%%%%%%%%%%%%%%%%%%%%%%%%%%%%%%%%%
\subsubsection{1+3 formalism}
%%%%%%%%%%%%%%%%%%%%%%%%%%%%%%%%%%%%%%%%%%%

In the 1+3 covariant approach~\cite{Ellis},  in which one introduces the worldline tangent vector $u^a$ ($u^a u_a=-1$),
one first introduces the projection tensor $h^a{}_b\equiv \delta^a{}_b+u^a u_b$.  This projection tensor defines two derivatives for any tensor ${T}^{a}{}_{b} $,  the covariant time derivative along the fundamental worldlines
\begin{equation}
\dot{T}^{a}{}_{b}\equiv  u^{e}\nabla_{e}T^{a}{}_{b}\,,\label{dot-def}
\end{equation} 
and the fully projected covariant derivative $D_e$ via
\begin{equation}
D_eT^{a}{}_{b} \equiv  h^a{}_f\,h^g{}_b\,h^r{}_e\nabla_r\,T^{f}{}_{g} \label{D-def}\ ,
\end{equation}
where we fully project on all free indices. 

The projection tensor $h_{ab}$ allows for  any 4-vector to be split into a scalar part parallel to $u^a$ and a 3-vector part orthogonal to $u^a$. Similarly, any second rank tensor may be covariantly and irreducibly split into scalar, vector and projected, symmetric, trace-free (PSTF) 3-tensor parts.

We can also split the covariant derivative of $u_a$ into its
irreducible parts as
\begin{eqnarray}
\label{eq:kin}
\nabla_{a}u_{b} &=& -\,u_a\,\dot{u}_b + D_{a}u_{b} \\
&=& -\,u_a\,\mathcal{A}_b + {\sfrac13}\,\Theta\,h_{ab} + \sigma_{ab}
+ \omega_{ab} \,.
\end{eqnarray}

This uniquely defines the following kinematic quantities: $\mathcal{A}_b=\dot{u}_b$ is the acceleration, 
the trace $\Theta = D_au^a{}$ is the {\em (volume) rate
of expansion\/} of the fluid (with $H = \Theta/3$ the Hubble
parameter), $\sigma_{ab} = D_{\langle a}u_{b \rangle}$ is the trace-free
symmetric {\em rate of shear\/} tensor describing the rate of
distortion of the matter flow, and $\omega_{ab} = D_{[a}u_{b]}$ is
the skew-symmetric {\em vorticity\/} tensor,
describing the rotation of the matter relative to a non-rotating frame. It is also useful define a scale factor $a$ along the fundamental worldlines via 
\begin{equation}
\frac{\dot{a}}{a} = \frac{1}{3}\,\Theta \,.
\end{equation}

%%%%%%%%%%%%%%%%%%%%%%%%%%%%%%%%%%%%%%%%%%%
\subsubsection{1+1+2 formalism}
%%%%%%%%%%%%%%%%%%%%%%%%%%%%%%%%%%%%%%%%%%%

We now employ the 1+1+2 formalism~\cite{CB}, that builds on the 1+3 formalism by allowing 
a further ``spatial" slicing with respect to a spacelike unit vector field $n^{a}$, 
which is orthogonal to the timelike 4-velocity vector ($u^{a}n_{a}=0$). The 1+3 projection tensor ${h_{a}}^{b}$  
combined with $n^{a}$ gives rise to a new projection 
tensor ${N_{a}}^{b}$,
which projects vectors and tensors onto the 2-surfaces orthogonal to $n^{a}$ and $u^{a}$ that are referred to as the `sheets':
\begin{eqnarray} 
N^a{}_b\equiv h^a{}_b-n^an_b=g^a{}_b+u^au_b-n^an_b\,.
\end{eqnarray}Analogously  to the 1+3 formalism,  $N_{a}{}^{b}$ defines two new derivatives for any object $T_{\cdots}{}^{\cdots}$: 
\begin{eqnarray}
\left({T}^{a}{}_{b}\right) '&\equiv&  n^{e}D_{e}T^{a}{}_{b}\,,\label{hatdef}\\
d_eT^{a}{}_{b} &\equiv &  N^a{}_f\,N^g{}_b\,N^r{}_e D_r\,T_{f}{}^{g} \label{deltadef}\,.
\end{eqnarray}
The $'$-derivative is the (spatial) derivative along the vector field $n^a$ in the surfaces orthogonal to $u^a$ analogous to the time derivative defined  in (\ref{dot-def}), and the $d_e$-derivative is the projected derivative on the sheet analogous to the projected derivative $D_e$ defined in (\ref{D-def}) \footnote{Note that the ${}'$-derivative was denoted by a hat ($\hat{}$), and the $d_e$ was denoted $\delta_e$ in Ref.~\cite{CB}.}.
See Ref.~\cite{CB} for a detailed presentation of the 1+1+2 formalism and the appendix for a summary of the key equations. 

%%%%%%%%%%%%%%%%%%%%%%%%%%%%%%%%%%%%%%%%%%%
\subsubsection{The LTB case}
%%%%%%%%%%%%%%%%%%%%%%%%%%%%%%%%%%%%%%%%%%%

\begin{figure}[htp]
\includegraphics[width=8cm]{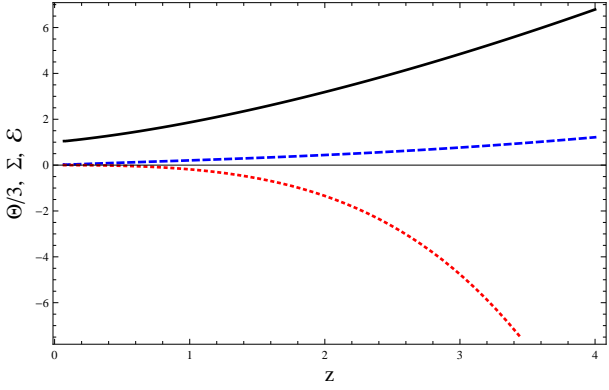}
\caption{Evolution of the expansion $\Theta$ (black, solid line),
the shear $\Sigma$  (blue dashed line)  and the electric part
of the Weyl tensor ${\cal E}$ (red dotted line) on the past light-cone
for the LTB model defined in Section~\ref{subsecIIc}.} 
\label{fig2}
\end{figure}

For a LTB spacetime all vectors and tensors as well as vorticity, acceleration and the magnetic part of the Weyl tensor vanish at background 
level. It can be further shown that the expansion, the only non-vanishing
component of the shear $\Sigma=n^\mu n^\nu\sigma_{\mu\nu}$
and of the electric part of the Weyl tensor ${\cal E}=n^\mu n^\nu E_{\mu\nu}$
are given by
\begin{equation}
 \Theta = 2H_\perp + H_\parallel, \quad
 \Sigma = \frac{2}{3}\left(H_\perp-H_\parallel\right), 
\end{equation}
and
\begin{equation}
 {\cal E}=\frac{8\pi G\rho}{3}-\frac{M}{R^3}.
\end{equation}
We depict their evolution on the past light-cone in Fig.~\ref{fig2},
and they satisfy the following general evolution equations:
\begin{eqnarray}
\dot\rho &=& -\Theta\rho,\ \label{eqnB1}\\
 \dot\Theta &=&-\frac{1}{3}\Theta^2 -4\pi G\rho -\frac{3}{2}\Sigma^2,\\
 \dot\Sigma & =& -\left(\frac{2}{3}\Theta +\frac{1}{2}\Sigma \right)\Sigma -{\cal E},\\
 \dot{\cal E} &=& -\left(\Theta -\frac{3}{2}\Sigma\right){\cal E} - 4\pi G\rho\Sigma. \label{eqnB4}
\end{eqnarray}

%%%%%%%%%%%%%%%%%%%%%%%%%%%%%%%%%%%%%%%%%%%
\subsection{The density perturbation equations}\label{sec:density perts}
%%%%%%%%%%%%%%%%%%%%%%%%%%%%%%%%%%%%%%%%%%%

We can decompose the standard dimensionless normalized density gradient ${\cal D}_a\equiv a ~ h_{ab}\nabla^b \rho/{\rho}$ into a 
part  $\Delta_a$ that lies in the sheet and a part perpendicular to the sheet:
\begin{eqnarray}
{\cal D}_a= a \frac{{\rho}'}{\rho} n_a+ a\frac{ d_a \rho}{\rho} \label{eqnD2}\equiv \, a \frac{{\rho}'}{\rho}n_a + \Delta_a\,,
\end{eqnarray}
where we recall that $\rho'= n^bD_b\rho$ and $ d_a \rho=N^b{}_aD_b\rho$.  

It is clear that $\Delta_a$ is gauge-invariant as it vanishes in the background LTB model. To see more clearly what this variable physically represents, let us define an infinitesimal vector $\delta x_{\bar{a}}\equiv N^b{}_a \,\delta x_b$ connecting two neighboring points on the sheet. It follows that the density evaluated at two neighboring points are related by
\begin{eqnarray}
\rho(x_a+\delta x_{\bar{a}})=\rho(x_a)+d_a\rho\,\,\delta x^{\bar{a}}\,.
\end{eqnarray}
In this way we can define a gauge-invariant density contrast $\delta$:
\begin{eqnarray}
\delta\equiv\frac{\rho(x_a+\delta x_{\bar{a}})-\rho(x_a)}{\rho(x_a)}=\frac{1}{a}\Delta_a\delta x^{\bar{a}}\,.
\end{eqnarray}
Defining the auxiliary perturbation variables in the sheet
\begin{eqnarray}
\delta\Theta=\frac{1}{a}\Theta_a\delta x^{\bar{a}},\quad
\delta\Sigma=\frac{1}{a}\mathcal{T}_a\delta x^{\bar{a}},\quad
\delta{\cal E}=\frac{1}{a}\mathcal{S}_{a}\delta x^{\bar{a}},
\end{eqnarray}
where 
\begin{eqnarray}
\Theta_a=a  \,d_a\Theta\,,~  \mathcal{T}_a=a  \,d_a\Sigma\,,~\mathcal{S}_{a}= a \,d_a{\cal E}\,,
\end{eqnarray}
we can derive a set of 4 gauge invariant perturbation equations (neglecting the magnetic part of the Weyl tensor)
\begin{eqnarray}
\dot{\delta}&=&-\delta\Theta+\frac{\rho'}{\rho}C_{\bar a}\delta x^{\bar a}\label{pertset0}\\
\left({\delta\Theta}\right)\dot{}&=&-\frac{2}{3}\Theta\delta\Theta-4\pi G\rho\delta-3\Sigma\delta\Sigma\nonumber\\
&&-\Theta' C_{\bar a}\delta x^{\bar a} \\
\left({\delta\Sigma}\right)\dot{}&=&\left(-\frac{2}{3}\Theta-\Sigma\right)\delta\Sigma-\frac{2}{3}\Sigma\delta\Theta-\delta{\cal E}\nonumber\\
 &&-\Sigma'C_{\bar a}\delta x^{\bar a} \\
\left({\delta\cal E}\right)\dot{}&=&-\left(\Theta-\frac{3}{2}\Sigma\right)\delta{\cal E}+\left(\frac{3}{2}{\cal E}-4\pi G\rho\right)\delta\Sigma\nonumber\\
&&-{\cal E}\delta\Theta-4\pi G\rho\Sigma\delta-{\cal E}'C_{\bar a}\delta x^{\bar a}\,,\label{pertset}
\end{eqnarray}
where $C_{\bar a}=\Sigma_{\bar a}+\alpha_{\bar a}$.
The equation for the connecting vector $\delta x^{\bar a}$ is only needed to background order,
\begin{eqnarray}\label{pertset4}
\delta\dot{x}^{\bar{a}}=\frac{1}{2}\left(\frac{2}{3}\Theta-\Sigma\right)\delta x^{\bar a}\,.
\end{eqnarray}
The set of equations~(\ref{pertset0}-\ref{pertset4}) provides a complete description of the evolution of the gauge invariant density contrast in situations where the magnetic part of the Weyl tensor can be neglected. This `silent" approximation can be thought of as neglecting the coupling between density perturbations and gravitational waves in pressure-free models and has been widely studied in the context of non-linear gravitational collapse \cite{silent}.

The above equations can be solved iteratively as follows. Expanding the perturbation variables in the form $\delta X^i=\delta X_0+ f^i_{\bar a}\delta x^{\bar a}$,  (with $i=1\ldots4$ and the short-hand notation $\delta X^1=\delta$ etc.), we obtain a set of homogeneous equations for the leading part of the perturbations
\begin{eqnarray}
\dot{\delta}_0&=&-\delta\Theta_0\\
\left({\delta\Theta}\right)\dot{}_0&=&-\frac{2}{3}\Theta\delta\Theta_0-4\pi G\rho\delta-3\Sigma\delta\Sigma_0 \\
\left({\delta\Sigma}\right)\dot{}_0&=&\left(-\frac{2}{3}\Theta-\Sigma\right)\delta\Sigma_0-\frac{2}{3}\Sigma\delta\Theta_0-\delta{\cal E}_0\\
\left({\delta\cal E}\right)\dot{}_0&=&-\left(\Theta-\frac{3}{2}\Sigma\right)\delta{\cal E}_0+\left(\frac{3}{2}{\cal E}-4\pi G\rho\right)\delta\Sigma_0\nonumber\\
&&-{\cal E}\delta\Theta-4\pi G\rho\Sigma\delta_0\,,\label{pertset2} 
\end{eqnarray}
with the corrections evolving according to equations of the form
\begin{eqnarray}
\dot{f}^i_{\bar a}=-\frac{1}{2}\left(\frac{2}{3}\Theta-\Sigma\right)f^i_{\bar a}+\alpha^i C_{\bar a}\,
\end{eqnarray}
{\bf with $\alpha^i\equiv(\rho'/\rho,-\Theta',-\Sigma',-{\cal E}' $}).
The homogeneous equations can then {\bf be} combined to give a pair of {\bf coupled} second order equations
\begin{eqnarray}
\ddot{\delta}_0&+&\frac{2}{3}\Theta\dot{\delta}_0-4\pi G\rho\delta_0=3\Sigma\delta\Sigma_0\,,\\
\left({\delta\Sigma}\right)\ddot{}_0&+&(\frac{5}{3}\Theta-\frac{1}{2}\Sigma)\left({\delta\Sigma}\right)\dot{}_0 \\
&-&\left(\frac{20}{3}\pi G\rho+\frac{2}{3}\Theta\Sigma-\frac{4}{9}\Theta^2-\frac{1}{2}{\cal E}+5\Sigma^2\right)\delta\Sigma_0\nonumber\\
&=&-\left(\frac{4}{3}\Sigma^2+\frac{2}{3}{\cal E}+\frac{2}{9}\Sigma\Theta\right)\dot{\delta}_0+\frac{20}{3}\pi G\rho\delta_0\,.\nonumber
\end{eqnarray}
The first one is similar to the standard one in FL models, but with a source term proportional to the background and perturbed 
shear so that it is coupled to a second differential equation for $\delta\Sigma_0$. 

In the cases where the above approximations are not valid, one needs to consider the
{\bf considerably} more complex full set of 1+1+2 equations that is described in the appendix.

%%%%%%%%%%%%%%%%%%%%%%%%%%%%%%%%%%%%%%%%%%%
\section{Transfer functions and initial conditions}\label{sec3}
%%%%%%%%%%%%%%%%%%%%%%%%%%%%%%%%%%%%%%%%%%%
\subsection{Back to Friedmann-Lema\^{\i}tre spacetime}
%%%%%%%%%%%%%%%%%%%%%%%%%%%%%%%%%%%%%%%%%%%

Before investigating the growth of structure in a LTB universe, let us
recall the relations between the background dynamics
and the growth of large scale structure in a flat $\Lambda$CDM model.

Since the cosmological constant develops no perturbation, and since the growth
of the density perturbations of the pressureless CDM component is dictated
by general relativity, it implies that in the linear regime, the evolution equations
reduce to the two equations
\begin{equation}\label{FLgrowtha}
 \dot\delta = -\delta\Theta,\qquad
 \delta\dot\Theta = -2H_{FL}\delta\Theta-4\pi G\rho\delta\,,
\end{equation}
that is to
\begin{equation}\label{Dgrowth}
 \ddot D +2H\dot  D -4\pi G\rho_\mat D  = 0,
\end{equation}
where the CDM density contrast has been decomposed as
$\delta(\bx,t) = D_+(t)\varepsilon_+(\bx) + D_-(t)\varepsilon_-(\bx)$,
$\varepsilon_\pm$ encoding the initial conditions. 
This equation can be recast in terms of $a$ as 
time variable~\cite{rigidity0,pubook} as
\begin{equation}\label{evoDLCDM}
 D'' +\left(\frac{\dd\ln H}{\dd a} +\frac{3}{a} \right)D'
 = \frac{3}{2}\frac{\Omega_{\mat0}}{a^5}D.
\end{equation}
Since $D_-=H$ is a solution, the
growing mode is given by
\begin{eqnarray}\label{gmodeFL}
 D_+   &=& \frac{5}{2}\frac{H(z)}{H_0}\Omega_{\mat0}\int_z^\infty\frac{(1+z')\dd z'}{[H(z')/H_0]^3}.
\end{eqnarray}
In our particular case, this can be integrated analytically as
\begin{eqnarray}
 D_+(z) &\propto& {}_2F_1\left[1,\frac{1}{3};\frac{11}{6};-\left(\frac{1-\Omega_{\mat0}}{\Omega_{\mat0}} \right)\frac{1}{(1+z)^3}\right]
 \nonumber\\
  &&\times\left(\frac{1-\Omega_{\mat0}}{\Omega_{\mat0}} \right)^{1/3}\frac{1}{1+z}.
\end{eqnarray}

This implies that if $H(z)$ is known from background observations, such
as SNIa, then $D_+(z)$ is fixed and is thus not an independent quantity in this
framework. There is a rigidity between the expansion history of the background and the growth rate~\cite{rigidity0,rigidity1}. 
This has attracted attention since this offers a test of the $\Lambda$CDM model~\cite{grtestfl,grtestfleps}.
Moreover, Eq.~(\ref{FLgrowtha}) implies that $\vartheta\equiv\delta\Theta/H$ is related to the density
contrast by
\begin{equation}\label{deff}
 \vartheta = -f(a) \delta
\end{equation}
where the function $f$ can be parameterized as~\cite{lindera,linderb}
\begin{equation}\label{defg}
 f_+\equiv\frac{\dd\ln D_+}{\dd\ln a} = \Omega_\mat(a)^\gamma
\end{equation}
for the growing mode.
Then, if general relativity is not modified, 
the index $\gamma$ can be computed once $H(z)$, or
equivalently the dark energy equation of state, is known and it was 
shown~\cite{linder2,ws98}  that $\gamma=0.55$
in the case of a $\Lambda$CDM. Since $\gamma$
can be measured from galaxy redshift surveys~\cite{survey}, thanks to the
redshift distortion~\cite{zdist}, it was argued that the value {\bf of the parameter
$\gamma$} offers a test of general relativity~\cite{grtestfl}.

%%%%%%%%%%%%%%%%%%%%%%%%%%%%%%%%%%%%%%%%%%%%
\subsection{Structure of the LTB growth rate equations}
%%%%%%%%%%%%%%%%%%%%%%%%%%%%%%%%%%%%%%%%%%%%

In a LTB spacetime, and under the approximations discussed
in Section~\ref{sec2}, we need to solve a set of 4 differential equations that can
be recast as
\begin{equation}
 \dot X_i = M_{ij}(r,t) X_j,
\end{equation}
where $X_i=(\delta,\delta\Theta,\delta\Sigma,\delta{\cal E})$ and where the
matrix $M_{ij}$ depends only on $r$ and $t$ through the background
quantities $\Theta$, $\Sigma$, $\rho$ and ${\cal E}$. It follows that the general
solution is of the form 
\begin{equation}
 X_i(r,t,\theta,\varphi) = T_{ij}(r,t) X_j(r,t_{\rm init},\theta,\varphi)
\end{equation}
where the angular dependence stems only from the initial conditions so
that the transfer functions depends only on $r$ and $t$. When interested
in observations such as weak-lensing, one integrates along the light-cone
so that only $T_{ij}(z)=T_{ij}[r_*(z),t_*(z)]$ is actually needed (see Ref.~\cite{zibin}).

%%%%%%%%%%%%%%%%%%%%%%%%%%%%%%%%%%%%
\subsubsection{Integrating the perturbation equations}\label{sub1}
%%%%%%%%%%%%%%%%%%%%%%%%%%%%%%%%%%%%

\begin{figure}[htp]
\includegraphics[width=10cm]{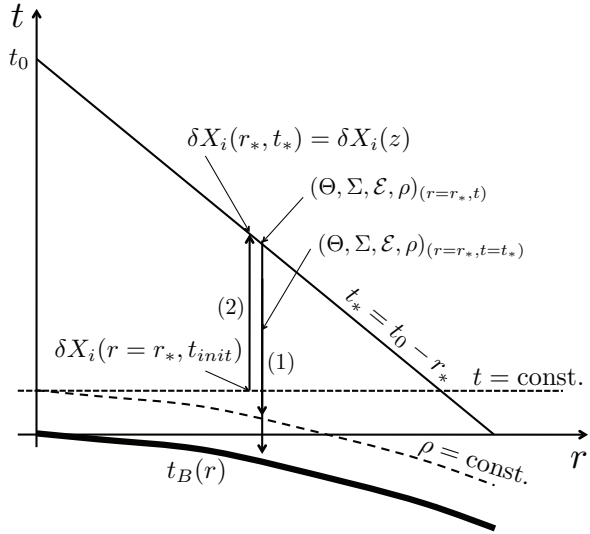}
\caption{Summary of the integration procedure. All quantities
are defined in the text. From the light-cone background quantities
one integrates the background equations toward the interior of the
light-cone at constant $r$ (1). Then, once an initial hypersurface
has been chosen, one integrates the perturbation equations
up to the light-cone (2).} 
\label{fig3b}
\end{figure}

In order to compute the transfer functions, we proceed as follows (see
Fig.~\ref{fig3b}).
\begin{itemize}
 \item Using the reconstruction along the past light-cone, we integrate for
 each $z$ the background equations (\ref{eqnB1}-\ref{eqnB4}) from $t=t_0-r_*(z)$
 toward the interior of the past light-cone at $r=r_*(z)$ constant.
 This provides the background quantities $\rho(t,r)$, $\Theta(t,r)$, $\Sigma(t,r)$,
 and ${\cal E}(t,r)$ for this particular $r$  in agreement
 with the constraint that $R=D_A$ and $\rho=\rho_{FL}$ on the past
 light-cone.
 \item We then solve the perturbation equations with this background functions
 from an initial time $t_{\rm init}$ up to $t_*(z)$ considering the four sets
 of initial conditions:
 \begin{equation}
  X_i^{(\alpha)}(r,t_{\rm init},\theta,\varphi) = \delta_i^\alpha, \qquad\alpha=1\ldots4.
 \end{equation}
 For each set $\alpha$, we obtain 4 transfer functions $T_{i\alpha}$.
\end{itemize}
Such an integration procedure is made possible because the background equations
involve no gradient term, so that each shell of constant $r$ evolves independently
and because,  in the silent approximation used in this work, the perturbation equations 
enjoy the same property. In this approximation, the spatial inhomogeneity
of the background spacetime reflects itself only on the fact that a structure
observed at a redshift $z$ had a growth history along the shell $r=r_*(z)$ which
is different from the other shell. This is a major difference with the FL situation
in which, as soon as we are dealing with a pressureless fluid,  
all the structures have the same growth rate from $z$ to 0, independently
of their growth rate at higher redshift.

\begin{figure}[htp]
\includegraphics[width=8cm]{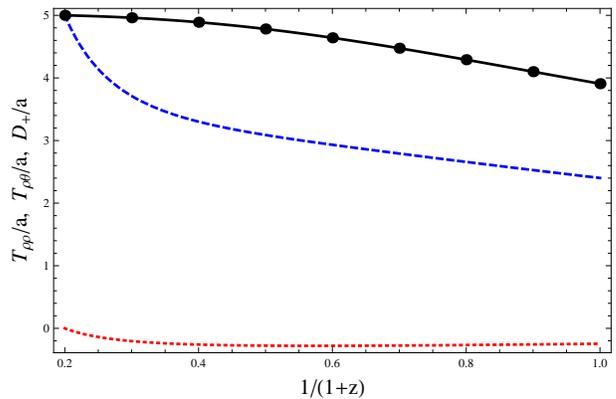}
\caption{The two transfer functions $T_{\rho\rho}$ (dashed blue line)   and $T_{\rho\theta}$
 (dotted green line) for a $\Lambda$CDM. We check that $T_{\rho\rho}-f_+(z_{\rm init})H(z_{\rm init})
T_{\rho\theta}$ (solid black line) is the growing mode $D_+$ (dots).} 
\label{fig5}
\end{figure}

First, this procedure can be tested on the FL model for which $\Sigma=0$ so
that $(\delta,\delta\Theta)$ decouples from $(\delta\Sigma,\delta{\cal E})$.
Focusing on the matter density contrast $\delta$, we compute the two
transfer functions $T_{\rho\rho}$ and $T_{\rho\theta}$ corresponding respectively to
the initial conditions $(\delta,\delta\Theta)=(1,0)$ and $(\delta,\delta\Theta)=(0,1)$. None
of these transfer functions corresponds to the growing mode (\ref{gmodeFL}) that is obtained from
the initial conditions  $(\delta,\delta\Theta)=(1,-f_+(t_{\rm init})H_{\rm init})$ so that
\begin{equation}
 D_+(z)= T_{\rho\rho}(z) - f_+(t_{\rm init})H_{\rm init}T_{\rho\theta}(z).
\end{equation}
Figure~\ref{fig5} shows that this is actually verified numerically. Note that even though we
reduce the dimension of the space of initial conditions by picking up the growing mode, we still need
the two transfer functions. Indeed one can also check that
$\delta\theta_+(z)= T_{\theta\rho}(z) - f_+(t_{\rm init})H_{\rm init}T_{\theta\theta}(z)=-Hf_+\delta_+$.

\subsubsection{Initial conditions}

To compute the transfer functions of the LTB model, we
need to specify an initial spatial hypersurface.
Any 3-dimensional spacelike hypersurface $\lbrace r = t_{\rm init}(r)\rbrace$ is
a priori possible but three choices can be argued to be natural: (i) a
constant time hypersurface, (ii) a constant density hypersurface
or (iii) a constant $t-t_B(r)$, i.e. a constant proper time after the
big bang. Indeed, in a FL model, these three possibilities reduce
to the same hypersurface.

\begin{figure}[htp]
\includegraphics[width=8cm]{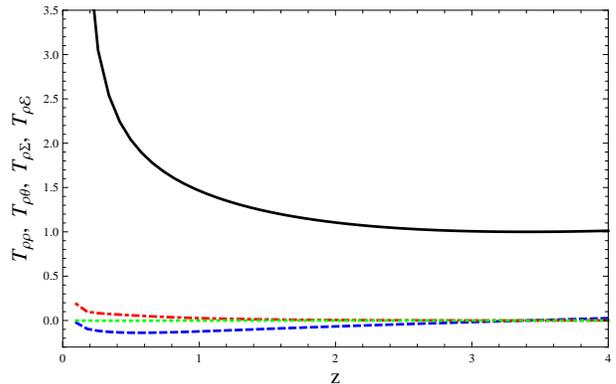}
\caption{The four transfer functions $T_{\rho\rho}$ (solid black line) and $T_{\rho\theta}$
(dashed blue line), $T_{\rho\sigma}$ (dot dashed red line) and 
$T_{\rho{\cal E}}$ (dotted green line)  for the LTB model.} 
\label{fig4}
\end{figure}

For the purpose of the illustration, we decide to set the initial
conditions on a constant time hypersurface, as in Ref.~\cite{zibin},
but we cannot justify this choice further here. Then, applying
the procedure described in Section~\ref{sub1}, we obtain the
transfer functions. Figure~\ref{fig4} describes the four
transfer functions needed to compute the density contrast
on the past light-cone.

%%%%%%%%%%%%%%%%%%%%%%%%%%%%%%%%%%%%%%%%
\subsubsection{Comparison of the two models}
%%%%%%%%%%%%%%%%%%%%%%%%%%%%%%%%%%%%%%%%

With the previous prescription, the two transfer
functions $T_{\rho\rho}$ and $T_{\rho\delta}$
for the two models look similar in shape and
amplitude.

To compare the growth rate, one needs to know the initial
conditions, or at least the relative initial value
of $\delta$, $\delta\Theta$, $\delta\Sigma$ and $\delta{\cal E}$.
As we have seen in Section~\ref{sub1}, a particular choice
of $(\delta,\delta\Theta)$ allows to pick up the
growing mode. We have no indication of the
linear combination of the transfer functions that
are related to the growing mode in the LTB case. In order
to compare the growth rate, we assume that at early time
the universe was well described by a FL-model and that
the void evolves at {\bf lower} redshift. In such a case, the initial
conditions can be set by their FL analog for the growing mode.
From Eqs.~(\ref{A33}) and~(\ref{A35}), we deduce that
\begin{equation}
 \frac{1}{2}\Theta\delta\Sigma_{\rm init} = \frac{2}{3}(8\pi G\delta\rho_{\rm init}),\quad
 \frac{1}{2}\delta{\cal E}_{\rm init} =-\frac{8\pi G}{3}\delta\rho_{\rm init}
\end{equation}
and for the growing mode
\begin{equation}
 \delta\Theta_{\rm init} = -f_+({\rm init})H({\rm init})\delta_{\rm init}.
\end{equation}
That would imply that $\delta = D_+^{\rm LTB}\delta_i$ with
\begin{eqnarray}\label{ltbD}
 D_+^{\rm LTB}(z)& =& T_{\rho\rho}(z) -f_+({\rm init})H({\rm init})T_{\rho\theta}(z)\\
 &+&(8\pi G\rho_{\rm init})\left[\frac{4}{9H_{\rm init}} T_{\rho\Sigma}(z)
  -\frac{2}{3} T_{\rho{\cal E}}(z)\right].\nonumber
\end{eqnarray}
It corresponds to the growing mode of the density perturbation if the universe has evolved from
a FL-phase and if the density perturbations had time to reach the FL growing mode before
the effect of the void on the evolution of the perturbations starts being non-negligible. Indeed,
we have no proof that it is the growing mode of the LTB-system. Note that
the 4 transfer functions are needed to described the evolution of the density, even though
the initial conditions can be reduced to the single random
variable $\delta_{\rm init}(r,t_{\rm init},\theta,\varphi)$.

\begin{figure}[htp]
\includegraphics[width=8cm]{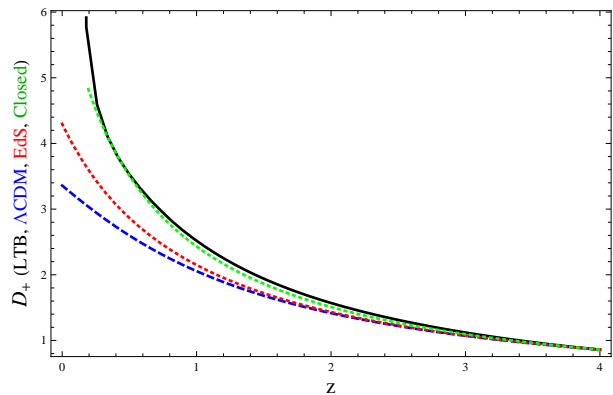}
\caption{Comparison of the density growing mode for a flat $\Lambda$CBM (blue dashed line)
an Einsein-de Sitter model (red dotted line) and our LTB model (black solid line) assuming the FL-like initial
conditions. It can be mimicked by a closed FL (green dotted line).} 
\label{fig6}
\end{figure}

Fig.~\ref{fig6} compares this solution to the growing mode of various FL-models. All
models are normalized at high redshift, and we can see that the LTB-model has more
structure at small redshift and that its growth rate is qualitatively similar
to the one of a closed FL-model.  We can also
note that it seems impossible to find a linear combination of the 4 transfer functions that
would mimic the growth rate of the $\Lambda$CDM model. This is an interesting conclusion,
given our ignorance of the initial conditions. Indeed, in general, we expect the initial
conditions to be not simply a linear combination but of the form
\begin{equation}
\delta\Theta_{\rm init} =f_\Theta(r)\delta_{\rm init}(r,\theta,\varphi),\quad
 \delta\Sigma_{\rm init} =f_\Sigma(r)\delta_{\rm init}(r,\theta,\varphi),
\end{equation}
and
\begin{equation}
 \delta{\cal E}_{\rm init} =f_{\cal E}(r)\delta_{\rm init}(r,\theta,\varphi),\qquad
\end{equation}
that is involving three arbitrary functions of $r$. Unless we have theoretical
constraints of $f_\Theta$, $f_\Sigma$ and $f_{\cal E}$, the
growth rate~(\ref{ltbD}) can be tuned at will. It follows that
the determination of the growth rate is more likely to teach us
on the initial conditions than to be used as an extra-data for the reconstruction
program.

As emphasized, in the $\Lambda$CDM, the growth rate and the background dynamics
are not independent so that we have relation such as Eq.~(\ref{defg}). For
the $\Lambda$CDM, $\gamma=0.55$ and slight deviation from this
value are expected for dark energy models in which general relativity is not
modified~\cite{grtestfl}. The index 
\begin{equation}
 \varepsilon(a) = \left[ \Omega_\mat(a) \right]^{-\gamma}\frac{\dd\ln D}{\dd\ln a} -1,
\end{equation}
introduced in Ref.~\cite{grtestfleps} should not deviate significantly from 0. $\gamma$ and
$\Omega_\mat$ are determined from background observations and thus coincide
with their $\Lambda$CDM values. It was shown that $\varepsilon$ can
typically vary between 0.05 and 0.25 for modification of general relativity
of the $f(R)$-class~\cite{grtestfleps}. Using the numerical solution corresponding
to Fig.~\ref{fig6}, we estimate that $\varepsilon$ can reach 0.1. The deviation
arises from the fact that the perturbation equations involve the shear and
the electric part of the Weyl tensor and indeed not from a
deviation from general relativity. This illustrates that it is indeed important
to ensure the validity of the Copernican principle when applying tests of
general relativity based on the large scale structures since they usually implicitly
assume its validity.

To finish, let us also emphasize that any observable, ${\cal O}$ say, entangles the properties
of the transfer functions and of the initial conditions. In general, it can be
written as an integral along the line of sight as
$$
 {\cal O}(z,n^a)=\int_0^z w(z,z')\delta[x^\mu(z',n^a)]\dd z',
$$
where $\delta$ stands for some combination of the perturbation variables and $w$ for
some window function that depends on the observable (e.g. for weak lensing
in a FL universe $\delta$ will correspond to twice the gravitational potential
and $w$ can be expressed in terms of angular distances). In a Friedmann-Lema\^{\i}tre
universe, as long as we focus on the evolution of dark matter, we have
that $\delta(x^i,t) = T(t)\delta_{\rm init}(x^i)$ so that the correlation function of ${\cal O}$
takes the form
\begin{eqnarray}
 &&\langle{\cal O}(z,n_1^a) {\cal O}(z,n_2^a)\rangle =
 \int_0^{z}\!\!\dd z' \int_0^{z}\!\!\dd z'' w(z,z')w(z,z'') \nonumber\\
 &&T(z')T(z'')\xi_{\rm init}(\sqrt{r^2(z')+r^2(z'') -2r(z')r(z'')\cos\theta} ),\nonumber
\end{eqnarray}
with $\cos\theta=n_1.n_2$ and 
where the correlation function of the initial condition, $\xi_{\rm init}$ is a function
of $\vert x^i_1-x^i_2\vert$ only because of isotropy and homogeneity. This integral
is usually easily evaluated in Fourier space. On small scales, the initial
conditions decouple from the evolution mainly because the power
is mainly carried by modes perpendicular to the line of sight (which is
at the basis of the Limber approximation) but the initial power
spectrum has to be such as this property holds, which is the
case for an almost scale invariant power spectrum. In a LTB
universe, there are two main differences with the FL case. First the transfer functions
depend on $r$ and $t$ so that  $\delta(x^i,t) = T(r,t)\delta_{\rm init}(x^i)$ 
and second the initial correlation dunction is a function 
$\xi_{\rm init}(r(z'),r(z''),\cos\theta )$
\begin{eqnarray}
 &&\langle{\cal O}(z,n_1^a) {\cal O}(z,n_2^a)\rangle =
 \int_0^{z}\!\!\dd z' \int_0^{z}\!\!\dd z'' w(z,z')w(z,z'') \nonumber\\
 &&\qquad T[r(z'),t(z')]T[r(z''),t(z'')]\xi_{\rm init}[r(z'),r(z''),\cos\theta].\nonumber
\end{eqnarray}
It follows that the spatial dependency of the transfer functions mixes with the
one of the initial conditions and that we cannot ensure a priori that a 
condition similar to the Limber property will exist.

Indeed, once a theory of the initial conditions compatible with the origin
of a large scale inhomogeneity exist, one can use our analysis to
compute the angular power spectrum of the observable ${\cal O}$. While
being focused on the properties of the transfer functions, our work however
show that if $\xi_{\rm initi}$ is identical as in a FL universe then the effect
of the evolution can lead to significant effect.

%%%%%%%%%%%%%%%%%%%%%%%%%%%%%%%%%%%%%%%%%%%
\section{Discussion}\label{sec4}
%%%%%%%%%%%%%%%%%%%%%%%%%%%%%%%%%%%%%%%%%%%

In this article, we have used a LTB model that
mimics a FL model on the past light-cone in order to shown
that these two models can still be distinguished by background observations
that encode information ``off'' the light-cone, as for instance
the time drift of cosmological redshift. While such an
observation was advocated as a test of the Copernican
principle in Ref.~\cite{uzanPC}, the amplitude of the
difference with the $\Lambda$CDM prediction had not
been estimated. We have shown that it can be significant and
that it can allow to exclude a large class of LTB models
even though they look similar to FL models at the background level.

Then, we investigated the information that can be
extracted from large scale structure. Assuming that
the curl of the magnetic part of the Weyl tensor can be neglected,
we have shown that one can extract a closed system
of 4 perturbation equations.  Our derivation  clarifies the link with
the gauge invariant variables of the $1+1+2$ formalism.
On small angular scales, we argued that one
can make a silent approximation. Under such conditions,
our set of equations for scalar perturbations reduce to the ones 
used in Ref.~\cite{zibin}.

We have detailed a procedure to compute the transfer functions
of this system of perturbation equation, once the background
data is known on the past light-cone and we explained how the corrections
at first order in the connecting vector can be obtained.
We emphasized the difficulty in determining a set of the initial conditions, i.e. to define
the hypersurface on which they are defined and the relation that
could exist between the different modes (in particular to 
extract the growing mode).

Indeed, our discussion of the perturbation dynamics is more illustrative 
than quantitative for two reasons. The first one is mainly technical since we have used
the silent approximation instead of solving the full set of 1+1+2
equations. This can only be achieved numerically and is left for further investigation.
The second is related to an intrinsic limitation concerning our ignorance
of the initial conditions: (1) we do not know the hypersurface on which
they have to be specified and (2) while we have been able to
compute the transfer function, we are not able to specify their
exact combination that corresponds to the growing mode. 
While in a FL universe, the late behaviour of the growth function
of dark matter reduces to a function of time, or equivalently
of redshift, it is in the LTB case a function of $t$ and $r$
that reduces to a function of $z$ on the light-cone. This 
implies a further limitation since the large scale structure
properties strongly entangle the evolution and the initial conditions. It
is thus difficult to close the reconstruction
program unless we have some constraints on the initial condition.
Nevertheless, our investigation shows that we cannot find
a linear combination (with constant coefficients) of the
transfer functions that reproduce the growing mode
of the $\Lambda$CDM. It also illustrates explicitely 
one limitation of the tests of general relativity
based on the large scale structure (e.g. the $\gamma$-index is difficult to generalize
to a LTB spactime because Eqs.~(\ref{deff}-\ref{defg}) have to be reconsidered).

This emphasizes the importance to test the Copernican
principle to validate the test of general relativity based
on the properties of large scale structures and illustrates
the effect of the assumption on the large scale geometry of our
universe on these tests.
In that respect both the time drift of the cosmological redshifts and the growth of
density perturbations give access to the spacetime structure
{\em beyond the light-cone} and are thus key observations for our
understanding of the geometry of our universe.

\begin{acknowledgements}
We thank C. Clarkson, G.F.R Ellis and Y. Mellier for useful discussions
and comments.
\end{acknowledgements}

%%%%%%%%%%%%%%%%%%%%%%%%%%%%%%%%%%%%%%%%%%%

%%%%%%%%%%%%%%%%%%%%%%%%%%%%%%%%%%%%%%%%%%%
\pagebreak
\begin{widetext}
\appendix
\section{Full set of perturbation equations in the 1+1+2 formalism}\label{appA}

If the magnetic part of the Weyl tensor or the source term $(\Sigma_a+\alpha_a)\delta^{\bar a}$ cannot be neglected, the linear perturbations around a LTB background are no longer described by equations~(\ref{pertset0}-\ref{pertset4}), and the complete set of first order equations below will successively be required to obtain a closed system of linear perturbations.  

Let us first recall that in the 1+1+2 framework any 3-vector or PSTF 3-tensor $T_{ab}$ can be irreducibly decomposed with respect to $n^{a}$ into scalar, vector and tensor parts:
\begin{equation}\label{tensor-decomp}
 T_{ab}={\mathcal T}\bra{\n_a\n_b-\:12\N_{ab}}+2{\mathcal T}_{(a}\n_{b)}+{\mathcal T}_{{ab}},
\end{equation}
where
\bea
{\mathcal T}&\equiv &\n^a\n^bT_{ab}=-\N^{ab}T_{ab},\nonumber\\
{\mathcal T}_a&\equiv &\N_a^{~b}\n^cT_{bc},\nonumber\\
{\mathcal T}_{ ab}&\equiv &\bra{\N_{(a}^{~~c}\N_{b)}^{~~d}-\:12\N_{ab}\N^{cd}}T_{cd}\label{PSTF-TT}.
\eea
The shear $\sigma_{ab}$ e.g. will be decomposed into the scalar part $\Sigma=\n^a\n^b\sigma_{ab}$, vector part $\Sigma_a= \N_a^{~b}\n^c\sigma_{bc}$ and tensor part $\Sigma_{ab}=\bra{\N_{(a}^{~~c}\N_{b)}^{~~d}-\:12\N_{ab}\N^{cd}}\sigma_{cd}$. 

We  may also decompose the covariant derivative of $\n^a$
orthogonal to $u^a$ in analogy to the 1+3 decomposition (\ref{eq:kin}):
\begin{equation}
D_a\n_b=\n_a a_b+\:12\phi \N_{ab}+\xi\lc_{ab}+\zeta_{ab},
\end{equation}
where
\bea
a_a &\equiv &\n^c D_c\n_a=\n'_a,\\
\phi &\equiv &d_a \n^a,\\
\xi &\equiv &\:12\lc^{ab}d_a\n_b,\\
\zeta_{ab} &\equiv &d_{\lb a}\n_{b\rb}.
\eea
Along the direction $n^a$, $\phi$ represents
the sheet expansion, $\zeta_{ab}$ is the shear of $\n^a$, $\hatn^a$ its acceleration, while $\xi$ represents a `twisting' of
the sheet.
For the LTB background, $\zeta_{ab},\, \n^a$ and $\xi$ are first order quantities. Furthermore, we restrict ourselves to perfect fluid perturbations where pressure, 4-acceleration and cosmological constant vanish at all orders.
%--------------------------------------------------
\subsection{Source evolution}
%--------------------------------------------------
The source evolution is then governed by the propagation equations
\bea
\dot{\Sigma}_a&=&-\left(\frac{2}{3}\Theta+\frac{1}{2}\Sigma\right){\Sigma}_a-\frac{3}{2}{\Sigma}{\alpha}_a-{\mathcal E}_a\\
\dot\E_{\bar a}+\:12\lc_{a}^b \H'_b &=& \frac{3}{4}\lc_{ab}d^b\H+\:12\lc_{bc}d^b\H^c_{~a}-\:12\bra{8\pi G\rho-\:32\E}\Sigma_a
-\:32{\E}\alpha_a
+\bra{\:34\Sigma-\Theta}\E_a-{\:14\phi}\lc_{ab}\H^b\\
\dot\Sigma_{\lb ab\rb}&=&-\bra{\:23\Theta+\:12\Sigma}\Sigma_{ab} 
    -\E_{ab} \\
\dot\E_{\lb ab\rb}-\lc_{c\lb a}\H'_{b\rb}{}^{c} &=&
    -\lc_{c\lb a}d^c\H_{b\rb} -\:12\bra{8\pi G\rho+3\E}\Sigma_{ab} -\bra{\Theta+\:32\Sigma}\E_{ab}+{\:12\phi}\lc_{c\lb a}\H_{b\rb}^{~~c}\\
\dot\H&=&-\lc_{ab}d^a\E^b-3\xi\E
    +\bra{\Theta+\:32\Sigma}\H\\
\dot\H_{\bar a}-\:12\lc_{a}{}^b\E'_b &=&
    -\:34\lc_{ab}d^b\E -\:12\lc_{bc}d^b\E^c_{~a}   +\:34{\E}\lc_{ab}a^b +{\:14\phi}\lc_{ab}\E^b+\bra{\:34\Sigma-\Theta}\H_a \\
\dot\H_{\lb ab\rb}+\lc_{c\lb a}\E'_{b\rb}{}^{c}&=& \lc_{c\lb a}d^c\E_{b\rb}
    +\:32{\E}\lc_{c\lb a}\zeta_{b\rb}^{~~c}-{\:12\phi}\lc_{c\lb a}\E_{b\rb}^{~~c}
    -\bra{\Theta+\:32\Sigma}\H_{ab}\\
\dot a_{\bar a}-\alpha'_{\bar a}&=&{\:12\phi}\alpha_a +\bra{\:13\Theta+\Sigma}{a_a}
    -{\:12\phi}{\Sigma_a}+\lc_{ab}\H^b\\
\dot\phi &=&-\bra{\:23\Theta-\Sigma}{\:12\phi}+d_a\dotn^a\\
\dot\xi &=& \bra{\:12\Sigma-\:13\Theta}\xi+\:12\lc_{ab}d^a\dotn^b +\:12\H,\label{dotxinl}\\
\dot\zeta_{\lb ab\rb}&=&\bra{\:12\Sigma-\:13\Theta}\zeta_{ab} {-\:12\phi}\Sigma_{ab} +d_{\lb
a}\alpha_{b\rb}  -\lc_{c\lb
a}\H_{b\rb}^{~~c}\,,
 \eea
 where we use curly brackets to denote the PSTF with respect to $n^a$ part of a tensor.
 The last equation may not be required up to first order, but just at background level, since it only seems to appear coupled to first order terms. Note that there is no propagation equation for $\alpha_a$.

%--------------------------------------------------
\subsection{Auxiliary First Order equations}
%--------------------------------------------------
Auxiliary first order propagation equations  along $u^a$ are given by
\bea
-\:13\dot\Theta-\dot\Sigma &=& \bra{\:13\Theta+\Sigma}^2+\:43 \pi G{\rho}+\E\\
 -\dot\Theta&=&\:13\Theta^2+\:32\Sigma^2+4\pi G{\rho}\\
 \dot\Sigma&=&-\bra{\:23\Theta+\:12\Sigma}\Sigma-\E\\
 \dot \rho&=&-\Theta\rho\\
\dot{\cal E}&=&\bra{\:32\Sigma-\theta}{\cal E}-4\pi G \rho\Sigma+\lc_{ab}d^a{\cal H}^b,\label{Edot}
\eea
and the first order propagation equations along $n^a$ are 
\bea
\phi' &=&-\:12\phi^2-\left(\:13\Theta+\Sigma\right)\left(\:23\Theta-\Sigma\right)-\:{16}3\pi G\rho-{\cal E}+d_a\hatn^a\\
\xi'&=&-\phi\xi+\:12\lc_{ab}d^aa^b\\
\zeta'_{\lb ab\rb}&=&-\phi\zeta_{ab}+
d_{\lb a}\hatn_{b\rb }+\bra{\:13\Theta+\Sigma}\Sigma_{ab}-{\cal E}_{ab}\\
\Sigma'-\:23\Theta'&=&-\:32\phi\Sigma-d_a\Sigma^a\label{A32}\\
\Sigma'_{\bar a}&=&\:12d_a\Sigma +\:23d_a\Theta-\:32\phi\Sigma_a -\:32\Sigma a_a-d^b\Sigma_{ab}\label{A33}\\
\Sigma'_{\lb ab\rb}&=&d_{\lb a}\Sigma_{b\rb} -\:12\phi\Sigma_{ab}+\:32\Sigma\zeta_{ab}-\lc_{c\lb a}\H_{b\rb}^{~~c}\\
\E'-\:83\pi G\rho'&=&-d_a\E^a -\:32\phi{\E}\\
\E'_{\bar a}&=& \:12d_a\E+\:83\pi Gd_a\rho-d^b\E_{ab}-\:32{\E}a_a-\:32\phi{\E_a}-\Sigma\lc_{ab}\H^b\label{A35}\\
\H'&=& -d_a\H^a-\:32\phi\H\\
\H'_{\bar a}&=& \:12d_a\H-d^b\H_{ab}-\:32{\E}\lc_{ab}\Sigma^b+\:32\Sigma\lc_{ab}\E^b -\:32\phi\H_a
\eea

%%%%%%%%%%%%%%%%%%%%%%%%%%%%%%%%%%%%%%
\section{Weak lensing for a central observer}
%%%%%%%%%%%%%%%%%%%%%%%%%%%%%%%%%%%%%%

One of the key observation to extract information about the growth rate of the
large scale structure is weak lensing, in particular using future tomographic survey.
We recall the Sachs equation~\cite{pubook,sachs,schneider}, 
which is the central equation governing gravitational lensing and then
consider it in a LTB universe.

%%%%%%%%%%%%%%%%%%%%%%%%%%%%%%%%%%%%%%%%%%%
\subsection{Sachs equation}
%%%%%%%%%%%%%%%%%%%%%%%%%%%%%%%%%%%%%%%%%%%

The tangent vector to a null geodesic is $k^a=\dd x^a/\dd\lambda$ and satisfied
$(1+z)= k^a u_a$ if we choose the value of the affine parameter $\lambda$
such that $k^a u_a=1$ today. It follows that it can be decomposed as
\begin{equation}
 k^a = -(1+z)\left(u^a + n^a \right).
\end{equation}
$n^a$ is the spatial direction of the photon and, in the particular
case in which the observer is at the center of the spherically symmetric
spacetime, the null geodesics are
radial and $n^a$  reduces to the radial vector used in the $1+1+2$ formalism.
We can then construct a basis by introducing two spatial unit vectors in the
2-dimensional sheet (i.e. the screen), $e_I^a$ with $I=1,2$ so that
$e_I^a e_{Ja}=\delta_{IJ}$ and $e_I^a  n_a=e_I^a u_a =0$.
It follows that
$h^a_b k^b = k^a + (1+z) u^a$, $h^a_b k^b k_a = (1+z)^2$ and 
$h^a_b e_{Ia}=  e_{Ib}$.
  
The central equation governing gravitational lensing is the Sachs
equation that derives from the geodesic deviation equation. Considering
a geodesic in the bundle $x^a=\bar x^a +\xi^a$, where the
vector $\xi^a$ can be decomposed as $\xi_0 k^a+
\sum_I \xi_I e_I^a$. The geodesic deviation equation then takes the form
$$
\frac{\dd^2 \xi_I}{\dd\lambda^2} = {\mathcal R}_I^J\xi_J
$$
where
${\mathcal R}_I^J\equiv {R^a}_{bcd}k^b k^c e_{Ia}e^{Jd}$.
The linearity of the geodesic equation implies that it is related to the initial value
of its derivative by a linear transformation $\xi_I(\lambda)={\mathcal D}_I^J (\dd\xi_J/\dd\lambda)_0$,
so that the Jacobi matrix ${\mathcal D}_I^J$ relates the shape of the cross-section of the
bundle to the basis $(e_I,e_J)$.
For a bundle converging at the observer, $\xi_I(0)=0$ and the matrix ${\mathcal D}_I^J$
evolves according to the Jacobi equation
\begin{equation}
 \frac{\dd^2}{\dd\lambda^2}{\mathcal D}_I^J = {\mathcal R}_I^K {\mathcal D}_K^J
\end{equation}
with ${\mathcal D}_I^J(0)=0$ and $(\dd{\mathcal D}_I^J/\dd\lambda)_0=\delta_I^J$.
Since the direction of observation is $\theta^I=(\dd\xi^I/\dd\lambda)_0$ and the
direction of the unlensed source $\theta_s^I=\xi^I(\lambda_s)/D_A(\lambda_s)$,
we conclude that the amplification matrix is ${\mathcal A}_I^J(\lambda)={\mathcal D}_I^J(\lambda)/D_A(\lambda)$.
This symmetric matrix is usually decomposed in terms of a shear $(\gamma_1,\gamma_2)$ and a convergence 
$\kappa$ as
\begin{equation}
 {\mathcal A}_{IJ}\equiv \left( 
\begin{array}{cc}
  1 - \kappa +\gamma_1 &  \gamma_2    \\
   \gamma_2 & 1-\kappa -\gamma_1        
\end{array}
\right).
\end{equation}
%-------------------------------------------------------
\subsection{The LTB case}
%-------------------------------------------------------
For an almost LTB universe, $P=0$, $q^a=0$, $\pi_{ab}=0$ so that $\dot u_a=0$.  It follows that the Riemann
tensor can be decomposed as ${R^{ab}}_{cd} = {_{\rm P}}{R^{ab}}_{cd} 
   + {_{\rm E}}{R^{ab}}_{cd} + {_{\rm H}}{R^{ab}}_{cd}$ with
\begin{eqnarray}
 {_{\rm P}}{R^{ab}}_{cd}  &=& \frac{16\pi G}{3}\rho
  \left(u^{[a}u_{[c}h^{b]}{}_{d]} + h^{a}{}_{[c}h^{b}{}_{d]}  \right) \\
 {_{\rm E}}{R^{ab}}_{cd}  &=& 4 u^{[a}u_{[c}E^{b]}{}_{d]}  +
   4h^{[a}{}_{[c}E^{b]}{}_{d]} \\
 {_{\rm H}}{R^{ab}}_{cd}  &=& 2\eta^{ab\epsilon}
 u_{[c}H_{d]\epsilon} +  2\eta_{cd\epsilon}
 u^{[a}H^{b]e}.
\end{eqnarray}
The two first terms contain background and first order term while the
third is purely first order. It follows that
\begin{eqnarray}
  (1+z)^{-2}{\mathcal R}_I^J&=&-\left[\left(4\pi G\rho+{\mathcal E}\right)\delta_I^J  +2E^a_b e_{Ia}e^{Jb}\right] -
  e_{Ia}e^{Jd}n_b\left(
  \eta^{abe}H_{de} + {\eta_{d}}^{be}H^a_e
  \right),
\end{eqnarray}
where we have used the decomposition of $k^a$.
Now, using the decomposition~(\ref{tensor-decomp}) for the electric and magnetic Weyl tensors
(reminding that $N_{ab}=h_{ab}-n_a n_b=e^1_a e^1_b +  e^2_a e^2_b$),
we obtain that
\begin{eqnarray}\label{eqRR}
 (1+z)^{-2}{{\mathcal R}_{IJ}}&=&-\left[4\pi G\rho\delta_{IJ}+2{\cal E}_a(e^a_I + e^a_J)  +  2{\cal E}_{ab}e^a_I e^b_J  \right] +
 n^b\eta_{abe}\left[{\cal H}e^a_{(I} e^e_{J)}-2e^a_{(I} e^d_{J)}{\cal H}_d^e
  \right].
\end{eqnarray}

At the background level,
only the first term contributes so that
\begin{eqnarray}
  {\mathcal R}_I^J&=&-4\pi G\rho(1+z)^2\delta_I^J
 \end{eqnarray}
and the Sachs equation reduces to
\begin{equation}
 \frac{\dd^2 }{\dd\lambda^2}{\mathcal D}_I^J = -4\pi G\rho(1+z)^2{\mathcal D}_I^J,
\end{equation}
from which we deduce that ${\mathcal D_I^J}=f(\lambda) I_I^J$. Since it is proportional to the 
identity matrix, it follows that the shear vanishes at the background level and we only have a convergence,
exactly as in the Friedmann case.
This was expected since for an observer at the center, the universe looks isotropic. The function $f$
then satisfies
$$
 \frac{\dd^2 }{\dd\lambda^2}f = -4\pi G\rho(1+z)^2f = -\frac{1}{2}R_{ab}k^a k^b f.
$$
The angular distance relates, by definition, the area of an object to the solid angle
under which it is observed, $\dd S^2 = D_A^2\dd\Omega^2$ satisfies the same equation as $f$
and has the same initial condition in 0 [$D_A(0)=0$ and $\dd D_A(0)/\dd\lambda=1$] so that $f=D_A$.
This follows from the fact that if ${\cal D}_{ab}$ is decomposed as
\begin{equation}
 {\mathcal D}_{IJ}\equiv \left( 
\begin{array}{cc}
  \hat{\theta} +\hat{\sigma}_1 & \hat{\sigma}_2 -\hat{\omega}   \\
   \hat{\sigma}_2+\hat{\omega} & \hat{\theta} -\hat{\sigma}_1        
\end{array}
\right),
\end{equation}
it can then be shown~\cite{sachs,Kantowski69,Pierlick04}
that the angular distance is related to the convergence by $\dd D_A/\dd\lambda =\theta D_A$. Now, derivating this
equation and expressing the derivative of the convergence in terms of the shear on gets~\cite{Pierlick04} that
$$
  \frac{\dd^2 }{\dd\lambda^2} D_A = -\left( \sigma_1^2 + \sigma_2^2 +  \frac{1}{2}R_{ab}k^a k^b\right)D_A,
$$
independently of the spacetime geometry.
Now, in the particular case in which $\sigma_{1/2}=0$, which is indeed the case at background level
since $\gamma_{1/2}=0$ we conclude that $D_A$ satisfies the same equation as $f$
and has the same initial condition in 0 [$D_A(0)=0$ and $\dd D_A(0)/\dd\lambda=1$] so that $f=D_A$.

Let us now turn to the perturbations. The expression~(\ref{eqRR}) shows that the perturbation will
enter both the shear and the convergence. Interestingly the shear vanishes at the background level.
We can formally integrate the Sachs equation to get
\begin{equation}\label{b12}
 {\cal A}_{IJ}=\int_0^\lambda\frac{D_A(\lambda')D_A(\lambda-\lambda')}{D_A(\lambda)}{\cal R}^{(1)}_{IJ}(\lambda')
 \dd \lambda'.
\end{equation}
Indeed the splitting ${\cal R}_{IJ}={\cal R}^{(0)}_{IJ}+{\cal R}^{(1)}_{IJ}$ is ambiguous but can be
used to obtain the expression of the shear since it vanishes at the background level. Note however
that the gauge issue strikes only the convergence since the shear vanishes at background level. We conclude
that, since $\gamma_1=({\cal A}_{11}-{\cal A}_{22})/2$ and $\gamma_2=({\cal A}_{12}+{\cal A}_{21})/2$ (where
the symmetrisation allows to get rid of the rotation that cannot be observed), we have
\begin{eqnarray}
 \frac{1}{2}({\cal R}_{11}+{\cal R}_{22}) &=& -4\pi G\rho -2{\cal E}_a(e^a_1+e_a^2) \nonumber \\
 \frac{1}{2}({\cal R}_{11}-{\cal R}_{22})   &=&  -2{\cal E}_a(e^a_1-e_a^2) -
 \sqrt{2}{\cal E}_{ab}P_+^{ab}\nonumber\\
 \frac{1}{2}({\cal R}_{12}+{\cal R}_{21}) &=& 
 -2{\cal E}_a(e^a_1+e_a^2) -
 \sqrt{2}{\cal E}_{ab}P_\times^{ab} -n^b\eta_{abe}{\cal H}_d^e P_\times^{ad},
 \nonumber
\end{eqnarray}
where
$$
 P_\lambda^{ab} = \frac{e_1^a e^1_b - e_2^a e^2_b}{\sqrt{2}}\delta_\lambda^+
 +
 \frac{e_1^a e^2_b + e_2^a e^1_b}{\sqrt{2}}\delta_\lambda^\times.
$$
We conclude that the shear is expressed only in terms of gauge invariant variables which
is  not the case of the convergence since it is non-vanishing at the background level. However
$\kappa(\lambda,n^a)$ so that at a fixed $\lambda$ (redshift), one can extract the
effect arising from the perturbation by taking a derivative in the sheet (since
the convergence induced by the background is isotropic). 

We recover the standard FL expression when ${\mathcal E}=0$ at
the background level. Here, the Sachs equation becomes
$$
\frac{\dd^2 }{\dd\lambda^2}{\mathcal D}_I^J = -\frac{3}{2}H_0^2\Omega_0(1+z)^5  {\mathcal D}_I^J.
$$
It can be shown~\cite{pubook,sb} that the solution of this equation is exactly $D_A(z)$ so that
${\mathcal D}_I^J =D_A(z)\delta_I^J$ (but the general argument~\cite{Pierlick04} ensures that it was expected).
At perturbation level, $E_{ab}= D_a D_b\Phi
-\frac{1}{3}\Delta\Phi h_{ab}= (\partial_a \partial_b\Phi
-\frac{1}{3}\Delta\Phi \delta_{ab})/a^2$ so that
\begin{eqnarray}
  \delta{\mathcal R}_{IJ}&=&-(1+z)^2\left[\left(4\pi G\delta\rho+\partial_{33}\Phi-\Delta\Phi\right)\delta_{IJ}
  +2D_ID_J\Phi\right]\nonumber\\ &=& -(1+z)^2\left[\partial_{33}\Phi\delta_{IJ} +2D_ID_J\Phi\right],\nonumber
\end{eqnarray}
once the Poisson equation is used. 

The expression ~(\ref{b12}) gives the generalisation of the expression of the shear in terms
of the perturbation variables of the 1+1+2 formalism for an observer seating at the center of
a LTB universe.\\

\end{widetext}


\begin{thebibliography}{99}
%%%%%%%%%%%%%%%%%%%%%%%%%%%%%%%%%%%%%%%%%%%

\bibitem{cmbr} 
 D.N. Spergel {\it et al.} Astrophys. J. Suppl. {\bf 148}, 175 (2003).
 
\bibitem{sneIa}
 S. Perlmutter {\it et al.}, Astrophys. J. {\bf 517}, 565 (1999); 
 A.G. Riess {\it et al.}, Astron. J. {\bf 116}, 1009 (1998);  
 J.L. Tonry {\it et al.}, Astrophys. J.  {\bf 594}, 1 (2003); 
 R.A. Knop {\it et al.}, Astrophys. J. {\bf 598}, 102 (2003); 
 A.G. Riess {\it et al.} Astrophys. J. {\bf 607}, 665 (2004).

\bibitem{lss}
 M. Tegmark {\it et al.}, Phys. Rev. {\bf D 69}, 103501 (2004); 
 U. Seljak {\it et al.}, Phys. Rev. {\bf D 71}, 103515 (2005).
 
\bibitem{baryon}
 D.J. Eisenstein {\it et al.}, Astrophys. J. {\bf 633}, 560 (2005).

\bibitem{wl}
  L. Fu {\it et al.}, Astron. Astrophys. {\bf 479}, 9 (2008). 

\bibitem{uzanGRG06}  
 J.-P. Uzan,
 Gen. Rel. Grav.  {\bf39}, 307 (2007).

\bibitem{uzanCUP}
 J.-P. Uzan,
 {\it Dark energy, gravitation and the Copernican principle},
 in {\it Dark energy}, P. Ruiz Lapuente Ed., (Cambridge University Press, 2010), 
 \url{arXiv:0912.5452}. 

\bibitem{uzanGRGrev}
 J.-P. Uzan,
 Gen. Rel. Grav.  (to appear), \url{arXiv:0908.2243}.  
 
\bibitem{demodels} 
 B. Ratra and P. J. E. Peebles, Rev. Mod. Phys. {\bf75}, 559 (2003).
 
\bibitem{uzanPC}
 J.-P. Uzan, C.C. Clarkson, and G.F.R. Ellis,
 Phys. Rev. Lett. {\bf100}, 191303 (2008).

\bibitem{stebbinsPC}
 J. Goodman, Phys. Rev. D {\bf52}, 1821 (1995);
 R.R. Caldwell and A. Stebbins,
 Phys. Rev. Lett. {\bf100}, 191302 (2008).
 
\bibitem{chrisPC}
 C. Clarkson, B. Bassett, and T. Lu,
 Phys. Rev. Lett. {\bf101}, 011301 (2008).

\bibitem{romanoPC}
 A.E. Romano, JCAP {\bf01}, 004 (2010);
 A.E. Romano, Phys. Rev. D {\bf76}, 103525 (2007).

\bibitem{ltb1}
 N. Mustapha {\it et al.},
 Mon. Not. R. Astron. Soc. {\bf292}, 817 (1997).
 
\bibitem{ltb2}
 M. C\'el\'erier, 
 Astron. Astrophys. {\bf353}, 63 (2000).
 
\bibitem{iguchi}
 H. Iguchi, T. Nakamura, and K.-I. Nakao,
 Prog. Theor. Phys. {\bf108}, 809 (2002).  

\bibitem{ltb3}
 M. C\'el\'erier {\it et al.},
 \url{arXic:0906.0905 [astro-ph.CO]}.
 
\bibitem{limitromano} 
 A.E. Romano, \url{0912.2866}.

\bibitem{ltb4}
 J. W. Moffat and D. C. Tatarski, Phys. Rev. D {\bf45}, 3512 (1992);
 {\it ibid.}, \url{arXiv:astro-ph/9404048};
 N. Sugiura, \etal, Phys. Rev. D {\bf60}, 103508 (1999);
 C. H. Chuang, J. A. Gu, and W.Y. Hwang, \url{astro-ph/0512651};
 D. Garfinkle, Class. Quant. Grav. {\bf23}, 4811 (2006); 
 D. J. Chung and A. E. Romano, Phys. Rev. D {\bf74}, 103507 (2006); 
 R. A. Vanderveld, \etal, Phys. Rev. D {\bf74}, 023506 (2006);
  H. Alnes, \etal, Phys. Rev. D {\bf73}, 083519 (2006);
  R. Mansouri, \url{astro-ph/0606703}; 
 M. N. C\'el\'erier, New Adv.  Phys. {\bf1}, 29 (2007);
 K. Enqvist and T. Mattsson, J. Cosmol. Astropart. Phys. {\bf02}, 019 (2007);
 C.-M. Yoo, \etal, Progr. Theor. Phys. {\bf120}, 937 (2008);
 J. P. Zibin, \etal, Phys. Rev. Lett. {\bf101}, 251303 (2008);
 J. Garcia-Bellido and T. Haugbolle, JCAP {\bf4}, 3 (2008);
 K. Bolejko and J. S. B. Wyithe, JCAP {\bf2}, 20 (2009).
 
\bibitem{ltbmetric}
 G. Lema\^{\i}tre, Ann. Soc. Sci. Bruxelles A {\bf53}, 51 (1933);
 R.C. Tolman, Proc. Nat. Acad. Sci. {\bf20}, 169 (1934);
 H. Bondi, Mon. Not. R. Astron. Soc. {\bf107}, 410 (1947).
 
\bibitem{stoegeretal}
 W. Stoeger, G. F. R. Ellis, and S. Nel, Class. Quant. Grav. {\bf9}, 509 (1992); 
 R. Maartens {\it et al.}, Class. Quant. Grav. {\bf13}, 253 (1996); 
 M. E. Araujo and W. R. Stoeger, Phys. Rev. D {\bf60}, 104020 (1999).
 
\bibitem{kolb}
 E.W. Kolb, and C.R. Lamb,
 \url{arXiv:0911.3852 [astro-ph.CO]}. 
 
\bibitem{chrisrevue}
C. Clarkson,
\url{arXiv:0911.2601 [astro-ph.CO]}. 

\bibitem{sandage}
 A. Sandage, Astrophys. J. {\bf136}, 319 (1962); 
 G. McVittie, Astrophys. J. {\bf136}, 334 (1962).
 
\bibitem{loeb}
 A. Loeb, Astrophys. J. {\bf499}, L111 (1998).

\bibitem{pasquini}
 L. Pasquini et al., The Messenger {\bf122}, 10 (2005).

\bibitem{ubm}
 J.-P. Uzan, F. Bernardeau, and Y. Mellier, 
 Phys. Rev. D {\bf77}, 021301(R) (2008).

\bibitem{liske}
 J. Liske {\em etal.},
 Mon. Not. R. Astron. Soc. {\bf386}, 1192 (2008). 
 
\bibitem{quartin}
 M. Quartin and A. Amensola,
 \url{arXiv:0909.4954}. 
 
\bibitem{corasaniti}
 P.-S. Corasaniti, D. Huterer, and A. Melchiorri, Phys. Rev. D {\bf75}, 062001 (2007); 
 A. Balbi and C. Quercellini, \url{arXiv:0704.235}; 
 H. Zhang et al., \url{arXiv:0705.4409}.
 
\bibitem{february} 
 S. February {\it et al.},
 \url{arXiv:0909.1479 [astro-ph.CO]}.
 
\bibitem{clarkson}
C. Clarkson, T.  Clifton and S. February
 JCAP 06 (2009) 025

\bibitem{zibin}
J.  P.  Zibin
Phys. Rev. D {\bf 78}, 043504 (2008). 

\bibitem{Ellis}
  G.F.R. Ellis, and H. van Elst,
``Cosmological Models", Carg\`{e}se Lectures 1998, 
 in {\it Theoretical and Observational Cosmology}, Ed. M Lachi\`eze-Rey, 
 (Dordrecht: Kluwer, 1999), 1. \url{arXiv:gr-qc/9812046};
 G.F.R.~Ellis, and M.~Bruni, Phys Rev D {\bf 40} 1804 (1989);  
 M.~Bruni,  P.K.S.~Dunsby, and G.F.R.~Ellis, Astrophys. J. {\bf 395} 34 (1992); 
 P.K.S.~Dunsby, M.~Bruni, and G.F.R.~Ellis, Astrophys.\ J.\  {\bf 395}, 54 (1992).

\bibitem{CB}
 C. A. Clarkson, R. K. Barrett, Class. Quant. Grav. {\bf20},  3855 (2003);
 C.  Clarkson {\it et al.}, Astrophys. J. {\bf 613}, 492 (2004);
 C. Clarkson,  Phys. Rev. D {\bf76}, 104034 (2007).
 
 \bibitem{rigidity0}
 T. Chiba, and T. Nakamura, Prog. Theor. Phys. {\bf 118}, 815 (2007);
 S. Nesseris, S., and L. Perivolaropoulos,  Phys. Rev. D {\bf77}, 023504 (2008).
 
 \bibitem{pubook}
  P. Peter, P., and J.-P. Uzan,
  {\it Primordial Cosmology}, Oxford University Press, 2009.  
  
 \bibitem{rigidity1} 
 E.V. Linder, Astropart. Phys. {\bf29}, 336 (2008);
 Y. Wang, JCAP {\bf0805}, 021 (2008).
 
\bibitem{grtestfl} 
  M. Ishak, M., A. Upadhye, and D.N. Spergel, Phys. Rev. D {\bf75}, 043513 (2006);
  Y. Wang {\it et al.}, Phys. Rev. D {\bf76}, 063503 (2007);
 M.J.  Mortonson, W. Hu, and D. Huterer, Phys. Rev. D {\bf79}, 023004 (2009);
 U. Alam, U., V. Sahni, and A.A. Starobinsky,  \url{arXiv:0812.2846}.

\bibitem{grtestfleps}  
 V. Acquaviva, {\it et al.}, Phys. Rev. D {\bf78}, 043514 (2008); 
 
 \bibitem{lindera}
 E.V.  Linder, and R.N. Cahn, Astropart. Phys. {\bf 28}, 481 (2007).

\bibitem{linderb}
 D. Huterer, and E.V. Linder, Phys. Rev. D {\bf 75}, 023519 (2007).
 
 \bibitem{linder2}
 E.V. Linder, Phys. Rev. D {\bf 72}, 043529 (2005).

\bibitem{ws98}
 L.M. Wang, and P.J. Steinhardt, Astrophys. J. {\bf 508}, 483 (1998).  
 
\bibitem{zdist}
 N. Kaiser, Month. Not. R. Astron. Soc. {\bf227}, 1 (1987);
 F. Bernardeau, {\it et al.}, Phys. Rept. {\bf367}, 1 (2002).
 
\bibitem{survey}
 M. Tegmark, A.J.S Hamilton,  and Y.  Xu, Month. Not. R. Astron. Soc. {\bf335}, 887 (2002);
 O. Le F\`evre, {\it et al.},  Astron. Astrophys. {\bf439}, 877 (2005);
 L. Guzzo, {\it et al.}, Nature {\bf451}, 541 (2008);
 A. Kosowsky, and S. Bhattacharya., \url{arXiv:0907.4202}. 

\bibitem{sachs}
 R.K. Sachs, Proc. R. Soc. London A {\bf264}, 309 (1961).

\bibitem{schneider}
 P. Schneider, J. Ehlers, and E.E. Falco, 
 {\it Gravitational Lenses} (Springer Verlag, Heidelberg, 1992).
 
\bibitem{Kantowski69} 
 R. Kantowski, J. Math. Phys., {\bf9}, 336 (1968).

\bibitem{Pierlick04} 
 V. Perlick, Living Rev. Relativity {\bf7}, 9 (2004).

\bibitem{sb}
 M. Bartelmann, and P. Schneider, Phys. Rept. {\bf340}, 291 (2001). 
 
 \bibitem{silent}
 M. Bruni, S. Matarrese, O. Pantano
Astrophys. J. {\bf445}, 958 (1995).

\end{thebibliography}
\end{document}